\documentclass[a4paper,11pt]{article}
\usepackage{amsfonts}
\usepackage{physics,amsmath}
\usepackage{amssymb}
\usepackage{mathtools}
\usepackage{colortbl}
\usepackage{xcolor}
\usepackage{cite}
\usepackage{hyperref}
\usepackage{mathrsfs}
\numberwithin{equation}{section}
\usepackage[left=2cm,right=2cm,top=3.5cm,bottom=3.5cm]{geometry}

\allowdisplaybreaks[1]
\hypersetup{
	colorlinks=true,
	linkcolor=ceruleanblue,
	filecolor=ceruleanblue,      
	urlcolor=ceruleanblue,
	citecolor=ceruleanblue,
}
\definecolor{ceruleanblue}{rgb}{0.0, 0.2, 0.6}
\newcommand{\vecb}[1]{\vectorbold*{#1}}

\date{\today}

\begin{document}

\begin{flushright} {\footnotesize YITP-23-138, IPMU23-0041}  \end{flushright}

\begin{center}
\LARGE{\bf Effective Field Theory of Black Hole Perturbations \\ 
in Vector-Tensor Gravity}
\\[1cm] 

\large{Katsuki Aoki$^{\,\rm a}$, Mohammad Ali Gorji$^{\,\rm b}$, Shinji Mukohyama$^{\,\rm a, \rm c}$, Kazufumi Takahashi$^{\,\rm a}$, and Vicharit Yingcharoenrat$^{\,\rm c}$}
\\[0.5cm]

\small{
\textit{$^{\rm a}$
Center for Gravitational Physics and Quantum Information, Yukawa Institute for Theoretical Physics, 
\\ Kyoto University, 606-8502, Kyoto, Japan}}
\vspace{.2cm}

\small{
\textit{$^{\rm b}$
Departament de F\'{i}sica Qu\`{a}ntica i Astrof\'{i}sica, Facultat de F\'{i}sica, Universitat de Barcelona, \\
Mart\'{i} i Franqu\`{e}s 1, 08028 Barcelona, Spain}}
\vspace{.2cm}

\small{
\textit{$^{\rm c}$
Kavli Institute for the Physics and Mathematics of the Universe (WPI), The University of Tokyo Institutes for Advanced Study (UTIAS), The University of Tokyo, Kashiwa, Chiba 277-8583, Japan}}
\vspace{.2cm}
\end{center}

\vspace{0.3cm} 

\begin{abstract}\normalsize
We formulate the effective field theory (EFT) of vector-tensor gravity for perturbations around an arbitrary background with a {\it timelike} vector profile, which can be applied to study black hole perturbations. The vector profile spontaneously breaks both the time diffeomorphism and the $U(1)$ symmetry, leaving their combination and the spatial diffeomorphism as the residual symmetries in the unitary gauge. We derive two sets of consistency relations which guarantee the residual symmetries of the EFT. Also, we provide the dictionary between our EFT coefficients and those of generalized Proca (GP) theories, which enables us to identify a simple subclass of the EFT that includes the GP theories as a special case. For this subclass, we consider the stealth Schwarzschild(-de Sitter) background solution with a constant temporal component of the vector field and study the decoupling limit of the longitudinal mode of the vector field, explicitly showing that the strong coupling problem arises due to vanishing sound speeds. This is in sharp contrast to the case of gauged ghost condensate, in which perturbations are weakly coupled thanks to certain higher-derivative terms, i.e., the scordatura terms. This implies that, in order to consistently describe this type of stealth solutions within the EFT, the scordatura terms must necessarily be taken into account in addition to those already included in the simple subclass. 
\end{abstract}

\vspace{0.3cm} 

\vspace{2cm}

\newpage
{
\hypersetup{linkcolor=black}
\tableofcontents
}

\flushbottom

\vspace{1cm}


\section{Introduction}

General relativity is a very successful theory of gravity that has been verified by many experimental tests including the recent one through the gravitational wave observations~\cite{LIGOScientific:2017vwq,LIGOScientific:2017zic}. Nevertheless, there is no a priori reason to assume that it is valid at any scales. For instance, astronomical and cosmological observations indicate the existence of dark energy that drives the late-time cosmic acceleration~\cite{SupernovaSearchTeam:1998fmf,SupernovaCosmologyProject:1998vns} (see \cite{Peebles:2002gy,Copeland:2006wr} for reviews). In the standard model of cosmology, cosmological constant plays the role of dark energy leading to the so-called cosmological constant problem that remains as an unsolved issue~\cite{Weinberg:1988cp,Peebles:2002gy,Padmanabhan:2002ji}. The existence of dark energy suggests possible modification of general relativity at large scales~\cite{Carroll:2003wy,Copeland:2006wr}. On the other hand, at very small scales and high-energy regimes, quantum gravity effects are expected to play important roles; therefore, the deviation from general relativity seems to be natural. This, for instance, may happen in the very early Universe around the big bang singularity and around the singularity of black holes. In this regard, modification of gravity at infrared and ultraviolet regimes seems to be reasonable.

There are many modified theories of gravity and one of the common features among most of them is the appearance of extra degree(s) of freedom.\footnote{See however \cite{Afshordi:2006ad,Lin:2017oow,Aoki:2018zcv,Iyonaga:2018vnu,Mukohyama:2019unx,Gao:2019twq,DeFelice:2020eju,DeFelice:2022uxv} for models of minimally modified gravity where no extra degree of freedom shows up.} The simplest possibility is when one scalar degree of freedom shows up. These theories, known as scalar-tensor theories, have been systematically formulated and studied in recent years. Starting from the pioneering work by Horndeski~\cite{Horndeski:1974wa}, the framework has been extended to the beyond Horndeski theories~\cite{Gleyzes:2014dya} and the degenerate higher-order scalar-tensor (DHOST) theories~\cite{Langlois:2015cwa,Crisostomi:2016czh,BenAchour:2016fzp,Takahashi:2017pje,Langlois:2018jdg} (see \cite{Langlois:2018dxi,Kobayashi:2019hrl} for reviews) and U-DHOST theories~\cite{DeFelice:2018ewo,DeFelice:2021hps}. Further extensions have been developed in \cite{Takahashi:2021ttd,Takahashi:2022mew,Takahashi:2023jro,Takahashi:2023vva} by use of a higher-derivative generalization of the disformal transformation.
The other simplest possibility is the vector-tensor theories, which include an extra vector field on top of the metric. If the vector field is massive, these theories can be thought of as an extension of scalar-tensor theories such that the longitudinal degree of freedom of the vector field plays the role of a shift-symmetric scalar field. This possibility has also been systematically studied in the context of generalized Proca (GP) theories~\cite{Tasinato:2014eka,Heisenberg:2014rta,Allys:2015sht,BeltranJimenez:2016rff,Allys:2016jaq} and their extensions~\cite{Heisenberg:2016eld,deRham:2020yet,Kimura:2016rzw}.

Which of scalar-tensor or vector-tensor theories are favored by nature can be only decided by the different observations like astronomical, cosmological, and gravitational wave observations. In order to test these theories with observations, in principle, one needs to explore all the parameter space in these theories which is very large and, hence, not an easy task to do. However, if we specify the setup for a particular purpose, the parameter space can be significantly reduced. One systematic way of doing so is to implement effective field theory (EFT) methods. For instance, in the case of cosmology, the background is given by the Friedmann-Lema\^{i}tre-Robertson-Walker (FLRW) metric. Due to the high symmetry of the cosmological background, the relevant parameter space for both the scalar-tensor and vector-tensor theories significantly reduces~\cite{Arkani-Hamed:2003pdi,Arkani-Hamed:2003juy,Cheung:2007st,Creminelli:2008wc,Gubitosi:2012hu,Bloomfield:2012ff,Gleyzes:2013ooa,Gleyzes:2014rba,Lagos:2016wyv,Aoki:2021wew,Aoki:2022ipw, Aoki:2022ylc}. Then, based on the universal description provided by the EFT method, model-independent predictions for the cosmological observables are possible. This is a very intriguing feature that shows the unique and powerful role of EFT in a better understanding of our Universe. 

The FLRW background can only describe our Universe at very large scales, i.e., scales larger than $100\,\mbox{Mpc}$ where it looks very homogeneous and isotropic. However, this is no longer true at smaller scales where there exist astronomical objects like galaxies, stars, and black holes. It is then very important and necessary to look for the formulation that describe not only cosmological background at large scales but also astronomical objects at small scales within the regime of validity of the EFT. The first step in this direction has been taken in \cite{Mukohyama:2022enj} where the EFT of scalar-tensor theories with a {\it timelike scalar field} for {\it any background} has been developed (see \cite{Khoury:2022zor,Mukohyama:2022skk} for the application of the setup to the black hole linear perturbations). While the EFT of vector-tensor theories with a {\it timelike vector field} for the cosmological background has been developed very recently~\cite{Aoki:2021wew}, the extension for {\it any background} has not been formulated yet. This is the main aim of this paper. 

The rest of the paper is organized as follows. In Section~\ref{sec2:EFT_general}, we systematically formulate the EFT of vector-tensor theories with a timelike vector field for perturbations around any background. In Section~\ref{sec:Dictionary}, we provide a dictionary between our EFT setup and the GP theories. In Section~\ref{sec4_BG_decoupling}, we find background equations for the case of a static and spherically symmetric background. We then find the corresponding quadratic action for the perturbations of the longitudinal mode at the decoupling limit. Finally, we summarize our results and conclude in Section~\ref{sec:conclusions}. Some technical analyses are presented in Appendix~\ref{app}.

\section{General construction of the EFT}\label{sec2:EFT_general}
\subsection{Symmetry breaking pattern and unitary gauge}
\label{sec:symmetry}
In this Section, we will explain how to construct the EFT action of vector-tensor theories with a timelike vector field on an arbitrary background metric. 
This kind of EFT will be particularly useful for describing the dynamics of black hole perturbations in the context of modified gravity theories and thus, e.g., determining their quasinormal mode spectrum.
As we have already mentioned, the EFT of vector-tensor theories on a cosmological background has been formulated in \cite{Aoki:2021wew}, to which the EFT we are going to formulate in this paper must reduce on a very large scale.

Similar to the EFT on a cosmological background~\cite{Aoki:2021wew}, our starting point is the existence of a non-vanishing timelike vector field~$v_\mu(x)$. This background vector field breaks part of the 4d spacetime diffeomorphism and defines a preferred threading of spacetime. In order to identify the symmetry breaking pattern of our EFT, it is more convenient to express the vector field as follows:
\begin{align}\label{v-def}
v_\mu = \partial_\mu{\tilde \tau} + g_M A_\mu \,;  \quad
g_M \equiv \frac{g}{M_\star} \;,
\end{align} 
where ${\tilde \tau}$ is a scalar field, $A_\mu$ is a vector field, $g$ is a gauge coupling, and $M_\star$ is a mass scale of the order of the Planck mass (which may be different from the observed Planck mass~$M_{\rm Pl}$). The vector field~\eqref{v-def} is invariant under the $U(1)$ transformation
\begin{align}\label{eq:combined_U(1)_time}
    {\tilde \tau} \rightarrow {\tilde \tau} - g_M \chi \;, \quad A_\mu \rightarrow A_\mu + \partial_\mu \chi \;,
\end{align}
where $\chi(x)$ is an arbitrary scalar function. One may choose the gauge $\tilde{\tau}=0$ in which (\ref{v-def}) is simply a rescaling of $v_{\mu}$. In our setup, on the other hand, the vector field~$v_{\mu}$ is assumed to have a timelike expectation value, and another gauge choice is more appropriate to understand the symmetry breaking pattern.
Since $v_{\mu}$ is timelike, it is convenient to choose a coordinate system such that ${\tilde \tau}$ coincides with the time coordinate~$\tau$. In this coordinate (or gauge) choice, the vector field~\eqref{v-def} takes the form
\begin{align}\label{preferred-vector}
  v_\mu\big|_{{\tilde \tau}=\tau} = \delta_{\ \mu}^\tau + g_M A_\mu \equiv \vectorbold*{\delta}_{\ \mu}^{\tau}\;.
\end{align} 
We call this gauge the {\it unitary gauge}. Now, as ${\tilde \tau}$ is identified with the time coordinate, the transformation~\eqref{eq:combined_U(1)_time} should be understood as the combination of the time diffeomorphism and the $U(1)$ transformation.
Therefore, our EFT setup possesses the residual $U(1)$ symmetry~\eqref{eq:combined_U(1)_time} on top of the usual 3d spatial diffeomorphism.\footnote{This symmetry breaking pattern is the same as that of gauged ghost condensation~\cite{Cheng:2006us}, as pointed out in \cite{Mukohyama:2006mm}.}

As it can be seen from \eqref{v-def} or \eqref{preferred-vector}, $v_\mu$ is not hypersurface-orthogonal in general and determines a {\it preferred time direction (or threading)}. On the other hand, in the EFT of scalar-tensor theories (see e.g., \cite{Cheung:2007st,Gubitosi:2012hu,Mukohyama:2022enj,Khoury:2022zor,Mukohyama:2022skk}), there is a {\it preferred time slicing}, defined by a hypersurface-orthogonal vector~$n_\mu \propto \partial_\mu {\tilde \tau}$ which is obviously not invariant under the residual $U(1)$ symmetry~\eqref{eq:combined_U(1)_time}.\footnote{Typically, the preferred time slicing in the EFT of scalar-tensor theories can be defined via $n_\mu\propto \partial_\mu \bar{\Phi}$, where $\bar{\Phi}$ is a time-dependent background of a scalar field.} Therefore, our EFT of vector-tensor theories with a timelike vector field is different than the usual EFT of scalar-tensor theories. Indeed, it is obtained from the EFT of shift-symmetric scalar-tensor theories \cite{Finelli:2018upr} when the shift symmetry is gauged~\cite{Cheng:2006us}.
Notice that, in the absence of the gauge coupling, $\vectorbold*{\delta}_{\ \mu}^{\tau}$ coincides with $\delta^\tau_{\ \mu}$, so that the EFT reduces to the one of scalar-tensor theories. Here and in what follows, bold quantities represent those with contributions of $A_\mu$. 

Note that here we have not assumed any specific forms of both the background metric and the background gauge field. Our assumptions so far are (i) the existence of a preferred timelike vector field~$\vectorbold*{\delta}_{\ \mu}^{\tau}=v_{\mu}|_{{\tilde \tau}=\tau}$ and (ii) the EFT in the unitary gauge is invariant under the residual $U(1)$ symmetry~\eqref{eq:combined_U(1)_time} in addition to the 3d spatial diffeomorphism.

It would be worth mentioning that it should be also possible to formulate the EFT with a preferred {\it spacelike} vector field by following a similar procedure presented below. However, we focus on the timelike case in the present paper, so that our EFT can smoothly connect to the EFT on a cosmological background on large scales.

\subsection{Building blocks}

The tensors which are invariant under the 4d diffeomorphism and original $U(1)$ symmetry like the metric~$g_{\mu\nu}$, 4d covariant derivative~$\nabla_\mu$, 4d Riemann tensor~$R_{\mu\nu\rho\sigma}$, and field strength of the vector field~$F_{\mu\nu}$ are trivial building blocks of our setup. In this Subsection, we look for the nontrivial building blocks which are invariant under the residual symmetries which consist of the 3d spatial diffeomorphism and the residual $U(1)$ symmetry~\eqref{eq:combined_U(1)_time}. The timelike vector~$\vectorbold*{\delta}_{\ \mu}^{\tau}$ is covariant under the 3d diffeomorphism and invariant under the residual $U(1)$ symmetry. Thus, its norm,
\begin{align}
\vecb{g}^{\tau\tau} \equiv  g^{\mu\nu} \vectorbold*{\delta}_{\ \mu}^\tau \vectorbold*{\delta}_{\ \nu}^\tau  = g^{\tau\tau} + 2 g_M A^\tau + g_M^2 A_\mu A^\mu \;,\label{eq:g_tautau}
\end{align}
which is negative for timelike $v_\mu$, is a building block. The unit future-directed timelike vector is defined by 
\begin{align}\label{eq:normal_tilde}
\vectorbold*{n}_\mu \equiv - \frac{\vecb{\delta}_{\ \mu}^\tau}{\sqrt{-\vectorbold*{g}^{\tau\tau}}} \;, \quad \vectorbold*{n}_\mu \vectorbold*{n}^\mu = -1 \;.
\end{align}
Then, it is natural to define the projection tensor, 
\begin{align}\label{eq:proj_hmunu}
\vectorbold*{h}_{\mu\nu} \equiv g_{\mu\nu} + \vectorbold*{n}_\mu \vectorbold*{n}_\nu \;,
\end{align}
which satisfies $\vectorbold*{h}^\mu_{\ \alpha} \vectorbold*{h}^\alpha_{\ \nu} = \vectorbold*{h}^\mu_{\ \nu}$ and $\vectorbold*{h}^\alpha_{\ \mu} \vectorbold*{n}_\alpha = 0$. 
The unit timelike vector and projection tensor allow one to define the quantities that are parallel or orthogonal to the vector~$\vectorbold*{\delta}_{\ \mu}^{\tau}$. However, there is no corresponding time slicing in general since $\vecb{n}_\mu$ is not hypersurface-orthogonal due to the non-vanishing vorticity which is provided by the transverse degrees of freedom in $A_\mu$. Therefore, the usual $3+1$ ADM decomposition is not suitable here. Nevertheless, one can use $1+3$ decomposition which takes into account the role of the vorticity. In Appendix~A of \cite{Aoki:2021wew}, the $1+3$ decomposition is introduced and some details of calculations are presented. Here we only present the results.

The first derivative of the unit preferred vector is characterized by the expansion tensor, the vorticity tensor, and the acceleration vector defined respectively as
\begin{align}
\vectorbold*{K}_{\mu\nu} &\equiv \vectorbold*{h}^\alpha_{\ (\mu|} \nabla_\alpha \vectorbold*{n}_{|\nu )}  \;, \label{eq:extrinsic}\\ 
\vectorbold*{\omega}_{\mu\nu} &\equiv \vectorbold*{h}^\alpha_{\ [ \mu|} \nabla_\alpha \vectorbold*{n}_{|\nu ] }  \;, \label{eq:omega} \\
\vectorbold*{a}_\mu &\equiv \vectorbold*{n}^\alpha \nabla_\alpha \vectorbold*{n}_\mu \;, \label{eq:a_mu}
\end{align}
where the symmetrization and the anti-symmetrization are denoted by $(\,)$ and $[\,]$, respectively, and $\nabla_\mu$ is the 4d covariant derivative. 

The 4d covariant derivative~$\nabla_\mu$ can be also decomposed into the orthogonal and parallel components respectively as
\begin{align}\label{covariantD-decomposition}
\pounds_{\vectorbold*{n}}\;, \quad
\vectorbold*{D}_\mu
\equiv
\vectorbold*{h}_\mu{}^\alpha \nabla_\alpha \;,
\end{align}
where $\pounds_{\vectorbold*{n}}$ and $\vectorbold*{D}_\mu$ denote the Lie derivative along ${\vectorbold*{n}}_\mu$ and orthogonal covariant derivative, respectively.

Using the projection tensor, it is straightforward to define the orthogonal spatial curvature from the commutator of the orthogonal covariant derivative~\cite{Aoki:2021wew}.\footnote{Note that there is no unique definition of the ``spatial curvature'' because of the absence of the spatial hypersurfaces, see Appendix~B of \cite{Aoki:2022ipw} and references therein.} In practice, we can compute the orthogonal spatial curvature through the Gauss equation:
\begin{align}
\vectorbold*{h}^{\alpha}{}_{\mu} \vectorbold*{h}^{\beta}{}_{\nu} \vectorbold*{h}^{\gamma}{}_{\rho} \vectorbold*{h}^{\delta}{}_{\sigma} R_{\alpha\beta\gamma\delta} =
{}^{(3)}\!\vectorbold*{R}_{\mu\nu\rho\sigma} + 2 \vectorbold*{W}_{[\rho|\mu} \vectorbold*{W}_{|\sigma]\nu}
\,, \label{Gauss-eq}
\end{align}
where $\vectorbold*{W}_{\mu\nu}\equiv \vectorbold*{K}_{\mu\nu}+\vectorbold*{\omega}_{\mu\nu}$ and $R_{\alpha\beta\gamma\delta}$ is the 4d Riemann tensor. The orthogonal spatial Ricci curvature and its trace can be defined in a usual manner: 
\begin{align}
{}^{(3)}\!\vectorbold*{R}_{\mu\nu} \equiv {}^{(3)}\!\vectorbold*{R}^\alpha_{\ (\mu|\alpha|\nu)} \;, \quad {}^{(3)}\!\vectorbold*{R} \equiv {}^{(3)}\!\vectorbold*{R}^\mu_{\ \mu} \;.
\end{align}
Note that the indices are symmetrized to define the orthogonal spatial Ricci curvature. While the anti-symmetric part of  ${}^{(3)}\!\vectorbold*{R}^\alpha_{\ \mu\alpha\nu}$ does not vanish differently from the case of the spatial curvature, it can be written in terms of $\vectorbold*{W}_{\mu\nu}$. Thus, it can be safely omitted from the building blocks of the EFT.

In the gauge field sector, the field strength tensor,
\begin{align}
F_{\mu\nu} = \partial_\mu A_\nu - \partial_\nu A_\mu \;,
\end{align}
can be decomposed into the electric part~$\vecb{E}_\mu$ and the magnetic part~$\vecb{B}_{\mu\nu}$ as
\begin{align}
F_{\mu\nu} = -2 \vectorbold*{E}_{[\mu} \vectorbold*{n}_{\nu ]} + \vectorbold*{B}_{\mu\nu} \;,
\end{align}
where 
\begin{align}\label{eq:B_munu}
\vectorbold*{E}_\mu \equiv \vectorbold*{n}^\alpha F_{\mu \alpha} \;, \quad \vectorbold*{B}_{\mu\nu} \equiv \vectorbold*{h}^\alpha_{\ \mu} \vectorbold*{h}^\beta_{\ \nu} F_{\alpha \beta} \;.
\end{align}
For later convenience, we introduce a rank-2 tensor~$\vecb{E}_{\mu\nu}$ via
\begin{align}
\vecb{E}_{\mu\nu} \equiv \epsilon_{\mu\nu}^{\ \ \ \rho \alpha}\vecb{n}_\rho \vecb{E}_\alpha \;. \label{eq:E_munu}
\end{align}
Here, $\epsilon_{\mu\nu\alpha\beta}= \sqrt{-g}\, \varepsilon_{\mu\nu\alpha\beta}$ denotes the covariant Levi-Civita tensor, 
where $\varepsilon_{\mu\nu\alpha\beta}$ is the totally anti-symmetric Levi-Civita
symbol with $\varepsilon_{0123} = 1$. 
Note that we have the relation $\vecb{E}_\mu = -\epsilon_{\mu}^{\ \ \nu \alpha \rho} \vecb{n}_{\nu} \vecb{E}_{\alpha \rho}/2$.
The reason why we have introduced this rank-2 tensor~$\vecb{E}_{\mu\nu}$ will become clear in Subsection~\ref{sec:expand_EFT}. 
From the definitions~(\ref{eq:B_munu}) and (\ref{eq:E_munu}), we explicitly see that the tensors~$\vecb{B}_{\mu\nu}$ and $\vecb{E}_{\mu\nu}$ are anti-symmetric with respect to the indices~$\mu$ and $\nu$.
Also, among the tensors~$F_{\mu\nu}$, $\vecb{B}_{\mu\nu}$, and $\vecb{E}_{\mu\nu}$, only two of them are independent, so throughout this paper we are going to only work with $\vecb{B}_{\mu\nu}$ and $\vecb{E}_{\mu\nu}$. 
Actually, it is useful to note that $F_{\mu\nu}$ can be expressed in terms of the tensors~$\vecb{B}_{\mu\nu}$ and $\vecb{E}_{\mu\nu}$ as
\begin{align}
F_{\mu\nu} = \vecb{n}_\lambda \vecb{E}_{\alpha \rho} \epsilon_{[\mu}^{\ \ \lambda \alpha \rho} \vecb{n}_{\nu ]} + \vecb{B}_{\mu\nu} \;, \quad F_{\mu\nu} F^{\mu\nu} = - 2 \vecb{E}^\mu \vecb{E}_\mu + \vecb{B}^{\mu\nu} \vecb{B}_{\mu\nu}  = \vecb{E}^\mu_{\ \nu} \vecb{E}^\nu_{\ \mu} - \vecb{B}^\mu_{\ \nu} \vecb{B}^\nu_{\ \mu} \;. 
\end{align}
Additionally, from Eqs.~(\ref{eq:omega}) and (\ref{eq:a_mu}), one finds that
\begin{align}
\vecb{\omega}_{\mu\nu} &=  - \frac{g_M}{2 \sqrt{-\vecb{g}^{\tau\tau}}} \vecb{B}_{\mu\nu} \;,  \label{eq:omega_2} \\
\vecb{a}_\mu &= \frac{g_M}{ 2 \sqrt{-\vecb{g}^{\tau\tau}}} \epsilon_{\mu}^{\ \ \nu \alpha \beta} \vecb{n}_{\nu}\vecb{E}_{\alpha\beta} - \frac{\vecb{h}^\nu_{\ \mu}\partial_\nu \vecb{g}^{\tau\tau}}{2 \vecb{g}^{\tau\tau}} \;.
\end{align}
The relations above imply that both $\vecb{a}_\mu$ and $\vecb{\omega}_{\mu\nu}$ are completely determined once $\vecb{E}_{\mu\nu}$, $\vecb{B}_{\mu\nu}$, $\vecb{h}_{\mu\nu}$, and $\vecb{g}^{\tau\tau}$ are specified. Thus, we can safely disregard the presence of both $\vecb{\omega}_{\mu\nu}$ and $\vecb{a}_\mu$ since they are not independent building blocks. Indeed, in practice, working with $\{\vecb{E}_{\mu\nu}, \vecb{B}_{\mu\nu}\}$ instead of $\{\vecb{a}_\mu,\vecb{\omega}_{\mu\nu}\}$ is better since the limit~$g_M\to0$ is manifest in this basis.

It is worth comparing the above $1+3$ decomposition with the usual $3+1$ ADM decomposition. In order to do so, we note that the normal vector~$\vecb{n}_\mu$ does not define a constant-$\tau$ hypersurface since there is a contribution coming from the transverse modes of $A_\mu$. Taking the limit~$g_M\to0$, we ignore the transverse modes and $\vecb{n}_\mu$ becomes hypersurface-orthogonal. Therefore, we should recover the usual well-known results of $3+1$ ADM decomposition in the limit~$g_M\to0$. In this limit, from \eqref{preferred-vector}, \eqref{eq:g_tautau}, and \eqref{eq:normal_tilde}, we see that $\vectorbold*{\delta}_{\ \mu}^{\tau} \to \delta_{\ \mu}^\tau$, $\vecb{g}^{\tau\tau} \to g^{\tau\tau}$, and $\vectorbold*{n}_\mu \to n_\mu = - {\delta}_{\ \mu}^\tau/\sqrt{-{g}^{\tau\tau}}$. Accordingly, the projection tensor~\eqref{eq:proj_hmunu} reduces to the induced metric of the spatial hypersurfaces~$\vectorbold*{h}_{\mu\nu} \to h_{\mu\nu} = g_{\mu\nu} + {n}_\mu {n}_\nu$. The expansion tensor reduces to the extrinsic curvature~$\vectorbold*{K}_{\mu\nu} \to K_{\mu\nu} = h^\alpha_{\ (\mu|} \nabla_\alpha n_{|\nu )}$ while, as expected, the vorticity tensor vanishes. Then, $\pounds_{\vectorbold*{n}}$ and $\vectorbold*{D}_\mu$ coincide with $\pounds_{{n}}$ and the 3d spatial derivative~${D}_\mu$  respectively and ${}^{(3)}\!\vectorbold*{R}_{\mu\nu\rho\sigma}$ reduces to the usual 3d curvature of the spatial hypersurfaces. It is interesting to note that in some particular situations, e.g., on a cosmological background, the geometrical quantities defined via $\vecb{n}_\mu$ and $\vecb{h}_{\mu\nu}$ coincide with the ones defined on the constant-$\tau$ hypersurface. Such a situation 
can be realized when the solution of $A_\mu$ is assumed to be irrotational, i.e., the transverse modes are absent and the vorticity tensor~$\vecb{\omega}_{\mu\nu}$ vanishes.

Now, we can easily write down the most general EFT action of vector-tensor theories that is invariant under the residual symmetries in the unitary gauge. On top of the 4d covariant building blocks~$g_{\mu\nu}$, $\nabla_\mu$, $R_{\mu\nu\rho\sigma}$, and $F_{\mu\nu}$, we have found independent building blocks~$\{\vecb{g}^{\tau\tau}, \vecb{K}_{\mu\nu}, \pounds_{\vecb{n}}, \vectorbold*{D}_\mu, {}^{(3)}\!\vectorbold*{R}_{\mu\nu\rho\sigma}, \vecb{B}_{\mu\nu}, \vecb{E}_{\mu\nu}\}$ which are covariant under the 3d diffeomorphism and invariant under the residual $U(1)$ symmetry~\eqref{eq:combined_U(1)_time}. As in the spatial curvature, the Weyl piece of ${}^{(3)}\!\vectorbold*{R}_{\mu\nu\rho\sigma}$ vanishes in $1+3$ dimensions and we only need to use the Ricci part~${}^{(3)}\!\vectorbold*{R}_{\mu\nu}$ as an independent building block. Therefore, the most general EFT action of vector-tensor theories in unitary gauge is given by
\begin{align}\label{eq:action_uni}
S = \int  \mathrm{d}^4x\sqrt{-g}~\mathcal{L}(\vectorbold*{g}^{\tau\tau}, {}^{(3)}\!\vectorbold*{R}_{\mu\nu}, \vectorbold*{K}_{\mu\nu}, \vectorbold*{E}_{\mu\nu}, \vectorbold*{B}_{\mu\nu}, \pounds_{\vecb{n}}, \vectorbold*{D}_\mu) \;,
\end{align}
where the contraction of indices, to construct scalar quantities, can be performed with respect to either the metric~$g_{\mu\nu}$ or the projection tensor~$\vecb{h}_{\mu\nu}$.
Note that here we use the orthogonal spatial Ricci curvature~${}^{(3)}\!\vecb{R}_{\mu\nu}$, rather than the 4d Ricci curvature~$R_{\mu\nu}$.\footnote{As discussed in \cite{Aoki:2021wew}, the 4d Ricci curvature~$R_{\mu\nu}$ contains time derivatives, which could result in an extra would-be ghostly state when one considers higher-order terms in the curvature. It was shown in \cite{Heisenberg:2014rta} that the action of the GP theories, which does not contain a ghost state, can be written in terms of ${}^{(3)}\!\vecb{R}_{\mu\nu}$.} It is important to emphasize that this EFT action in the unitary gauge can be generally applied to an arbitrary background geometry. 
However, this form of the EFT action is not practically useful to describe the dynamics of both backgrounds and perturbations. In the next Subsection, we will expand such a covariant form of the EFT in terms of perturbations of the building blocks. 

\subsection{Consistency relations: expansion around an inhomogeneous background}\label{sec:expand_EFT}
In this Subsection, following the procedure done in \cite{Aoki:2021wew,Mukohyama:2022enj}, we expand the general EFT action~\eqref{eq:action_uni} in the unitary gauge around a non-trivial $\tau$- and $\vec{x}$-dependent background. 
At the end, we will impose two sets of consistency relations among the EFT coefficients, which ensure the invariance of the EFT under both the 3d diffeomorphism and the residual $U(1)$ symmetry~\eqref{eq:combined_U(1)_time}. 

In what follows, for simplicity, we shall focus on lower-derivative terms by assuming that terms involving the derivative operators~$\pounds_{\vecb{n}}$ and $\vectorbold*{D}_\mu$ are suppressed. We then obtain a simplified version of the EFT action~\eqref{eq:action_uni} as follows:
\begin{align}\label{eq:action_uni2}
S = \int  \mathrm{d}^4x\sqrt{-g}~\mathcal{L}(\vectorbold*{g}^{\tau\tau},\vectorbold*{K}^\mu_{\ \nu},{}^{(3)}\!\vectorbold*{R}^\mu_{\ \nu},\vectorbold*{E}^\mu_{\ \nu},\vectorbold*{B}^\mu_{\ \nu}) \;.
\end{align}
Here, the tensors are all rank-2 and always have one upper index and one lower index.
Written in this form, the contraction of the indices is always done by the Kronecker delta~$\delta^\mu_{\ \nu}$, and hence the projection tensor~$\vecb{h}^\mu_{\ \nu}$ does not appear explicitly in the Lagrangian. 
This is why we have used the rank-2 tensor~$\vecb{E}^\mu_{\ \nu}$ rather than the vector~$\vecb{E}_\mu$.
If we regarded $\vecb{E}_\mu$ as an independent variable, then the Lagrangian would have terms like $\vecb{E}_\mu\vecb{E}^\mu=\vecb{h}^{\mu\nu}\vecb{E}_\mu\vecb{E}_\nu$, which explicitly contains the projection tensor. Note that the way of the contraction does not matter because the rank-2 tensors are either symmetric or anti-symmetric in their indices, e.g., $\vectorbold*{K}^{\mu}{}_{\nu} \vectorbold*{K}^{\nu}{}_{\mu}= \vectorbold*{K}^{\mu}{}_{\nu} \vectorbold*{K}_{\mu}{}^{\nu}$.

We define perturbations around an arbitrary background metric
 \begin{align}\label{eq:pert_geo_gauge}
 \begin{split}
 &\delta \vectorbold*{g}^{\tau\tau} \equiv \vecb{g}^{\tau\tau} - \bar{\vectorbold*{g}}^{\tau\tau}(\tau,\vec{x}) \;, \quad \delta \vectorbold*{K}^\mu_{\ \nu} \equiv \vectorbold*{K}^\mu_{\ \nu} - \bar{\vectorbold*{K}}^\mu_{\ \nu}(\tau, \vec{x}) \;, \quad \delta {}^{(3)}\!\vectorbold*{R}^\mu_{\ \nu} \equiv {}^{(3)}\!\vectorbold*{R}^\mu_{\ \nu} - {}^{(3)}\!\bar{\vectorbold*{R}}^\mu_{\ \nu}(\tau, \vec{x}) \;,
 \\
 &\delta \vectorbold*{E}^\mu_{\ \nu} \equiv \vectorbold*{E}^\mu_{\ \nu} - \bar{\vectorbold*{E}}^\mu_{\ \nu}(\tau,\vec{x}) \;, \quad \delta \vectorbold*{B}^\mu_{\ \nu} \equiv \vectorbold*{B}^\mu_{\ \nu} - \bar{\vectorbold*{B}}^\mu_{\ \nu}(\tau, \vec{x}) \;,
 \end{split}
 \end{align}
where the background values are denoted with a bar and are generally $\tau$- and $\vec{x}$-dependent. Note that all the background quantities depend on the background of $A_\mu$.
Since here we are interested in the EFT on an arbitrary background metric, the background for the spatial components of $A_\mu$ is allowed to be non-vanishing, which is in contrast to the case of cosmological background~\cite{Aoki:2021wew}.

One could separate $\vecb{K}^\mu_{\ \nu}$ and ${}^{(3)}\!\vecb{R}^\mu_{\ \nu}$ into the trace and traceless parts with respect to $\vecb{h}^\mu_{\ \nu}$, i.e.,
    \begin{align}
    \vecb{K}^\mu_{\ \nu}\equiv \frac{\vecb{K}}{3}\vecb{h}^\mu_{\ \nu}+\vecb{\sigma}^\mu_{\ \nu}\;, \quad
    {}^{(3)}\!\vecb{R}^\mu_{\ \nu}\equiv \frac{{}^{(3)}\!\vecb{R}}{3}\vecb{h}^\mu_{\ \nu}+\vecb{r}^\mu_{\ \nu}\;,
    \label{K_R3_decomposition}
    \end{align}
respectively.
This decomposition is especially useful on a cosmological background where the background values of $\vecb{\sigma}^\mu_{\ \nu}$ and $\vecb{r}^\mu_{\ \nu}$ vanish (see \cite{Aoki:2021wew}).
In any case, this is just a matter of notation and one could choose either $\{\vecb{K}^\mu_{\ \nu},{}^{(3)}\!\vecb{R}^\mu_{\ \nu}\}$ or $\{\vecb{\sigma}^\mu_{\ \nu},\vecb{r}^\mu_{\ \nu},\vecb{K},{}^{(3)}\!\vecb{R}\}$ as a part of the EFT building blocks.
We will discuss more on the relation between the two conventions in Appendix~\ref{app}.

Let us proceed with the action~\eqref{eq:action_uni2}.
The Taylor expansion of the action up to first order in perturbations defined in (\ref{eq:pert_geo_gauge}) is given by 
\begin{align}
S_{(1)} = \int  \mathrm{d}^4x \sqrt{-g}\bigg[\bar{\mathcal{L}} + \bar{\mathcal{L}}_{\vectorbold*{g}^{\tau\tau}}\delta \vectorbold*{g}^{\tau\tau}
+ \bar{\mathcal{L}}_{\vectorbold*{K}^\mu_{\ \nu}}\delta \vectorbold*{K}^\mu_{\ \nu}  + \bar{\mathcal{L}}_{{}^{(3)}\!\vectorbold*{R}^\mu_{\ \nu}}\delta {}^{(3)}\!\vectorbold*{R}^\mu_{\ \nu}  + \bar{\mathcal{L}}_{\vectorbold*{E}^\mu_{\ \nu}}\delta \vectorbold*{E}^\mu_{\ \nu} + \bar{\mathcal{L}}_{\vectorbold*{B}^\mu_{\ \nu}} \delta \vectorbold*{B}^\mu_{\ \nu} \bigg] \;, \label{eq:Taylor_first}
\end{align}
where $\bar{\mathcal{L}}_{X^\mu_{\ \nu}} \equiv (\partial \mathcal{L}/\partial X^\mu_{\ \nu})|_{\rm BG}$. The Taylor coefficients are made of the background quantities so they must have the same symmetries as the background if the background has. The first four terms in the action~(\ref{eq:Taylor_first}) were already present in the EFT of vector-tensor theories on the FLRW background~\cite{Aoki:2021wew}, whereas the last two terms with $\delta \vecb{E}^\mu_{\ \nu}$ and $\delta \vecb{B}^\mu_{\ \nu}$ are additionally present due to the fact that their coefficients~$\bar{\mathcal{L}}_{\vectorbold*{E}^\mu_{\ \nu}}$ and $\bar{\mathcal{L}}_{\vectorbold*{B}^\mu_{\ \nu}}$ may acquire non-vanishing values in a general background which is not necessarily isotropic.
As usual, the above action up to the first order in perturbations is responsible for the background dynamics of both the metric and the gauge field.  
Furthermore, the expansion of the action~(\ref{eq:action_uni2}) at second order gives
\begin{align}
S_{(2)} = \int  \mathrm{d}^4x &\sqrt{-g} \bigg[\frac{1}{2}\bar{\mathcal{L}}_{\vecb{g}^{\tau\tau}\vecb{g}^{\tau\tau}} (\delta \vecb{g}^{\tau\tau})^2  +  \bar{\mathcal{L}}_{\vecb{g}^{\tau\tau}\vecb{Z}^I{}^\mu_{\ \nu}} \delta \vecb{g}^{\tau\tau} \delta \vecb{Z}^I{}^\mu_{\ \nu}  +  \frac{1}{2}\bar{\mathcal{L}}_{\vecb{Z}^I{}^\mu_{\ \nu}\vecb{Z}^J{}^\alpha_{\ \beta}}\delta \vecb{Z}^I{}^\mu_{\ \nu} \delta \vecb{Z}^J{}^\alpha_{\ \beta}   \bigg] \;, \label{eq:Taylor_second}
\end{align}
where we have introduced a compact notation~$\vecb{Z}^I{}^\mu_{\ \nu} = \{\vecb{K}^\mu_{\ \nu}, {}^{(3)}\!\vecb{R}^\mu_{\ \nu}, \vecb{E}^\mu_{\ \nu}, \vecb{B}^\mu_{\ \nu}\}$ and defined $\bar{\mathcal{L}}_{\vecb{Z}^I{}^\mu_{\ \nu} \vecb{Z}^J{}^\alpha_{\ \beta}} \equiv (\partial^2 \mathcal{L}/\partial \vecb{Z}^I{}^\mu_{\ \nu} \partial \vecb{Z}^J{}^\alpha_{\ \beta})|_{\rm BG}$ etc.
Here and in what follows, we suppress the summation over the repeated indices~$I$ and $J$.

The time coordinate~$\tau$ and spatial coordinates~$\vec{x}$ are not the building blocks of the EFT and from the actions~(\ref{eq:Taylor_first}) and (\ref{eq:Taylor_second}), we see that both the residual symmetries of the EFT [i.e., the 3d diffeomorphism and the residual $U(1)$ symmetry~\eqref{eq:combined_U(1)_time}] are apparently broken due to the fact that the Taylor coefficients evaluated on the non-trivial background generally depend on both $\tau$ and $\vec{x}$. However, the fact that the whole Taylor series we have obtained is derived from the covariant action~(\ref{eq:action_uni}) which is invariant under the residual symmetries suggests that there must be a set of relations among the Taylor coefficients that guarantees invariance of the series expansions~(\ref{eq:Taylor_first}) and (\ref{eq:Taylor_second}) under the residual symmetries. 
Indeed, there are two sets of consistency relations: One is associated with the residual 3d diffeomorphism and the other with the residual $U(1)$ symmetry~\eqref{eq:combined_U(1)_time}. 
Note that the former is similar to the one obtained in the context of the EFT of scalar-tensor theories on a generic background~\cite{Mukohyama:2022enj}, while the latter is similar to the one derived in the EFT of vector-tensor theories on a cosmological background~\cite{Aoki:2021wew}. 

The first set of consistency relations which guarantees the 3d diffeomorphism invariance of the action is obtained by applying the chain rule with respect to the spatial derivatives. 
Note that we use the fact that the Taylor coefficients may implicitly depend on $\vec{x}$ through the background values. 
Then, the chain rule applied to the derivative of $\bar{\mathcal{L}}$ with respect to $\vec{x}$ gives
\begin{align}
\frac{\partial}{\partial \vec{x}} \bar{\mathcal{L}}(\tau,\vec{x})  = \ & \frac{\rm d}{{\rm d} \vec{x}} \bar{\mathcal{L}}(\vectorbold*{g}^{\tau\tau}, \vecb{K}^\mu_{\ \nu},{}^{(3)}\!\vecb{R}^\mu_{\ \nu},\vectorbold*{E}^\mu_{\ \nu}, \vectorbold*{B}^\mu_{\ \nu})  \bigg|_{\rm BG} \nonumber \\
= \  &\bar{\mathcal{L}}_{\vectorbold*{g}^{\tau\tau}}\frac{\partial \bar{\vectorbold*{g}}^{\tau\tau} }{\partial \vec{x}}
+ \bar{\mathcal{L}}_{\vectorbold*{K}^\mu_{\ \nu}}\frac{\partial \bar{\vectorbold*{K}}^\mu_{\ \nu}}{\partial \vec{x}}   + \bar{\mathcal{L}}_{{}^{(3)}\!\vectorbold*{R}^\mu_{\ \nu}}\frac{\partial{}^{(3)}\!\bar{\vectorbold*{R}}^\mu_{\ \nu}}{\partial \vec{x}}
+ \bar{\mathcal{L}}_{\vectorbold*{E}^\mu_{\ \nu}} \frac{\partial \bar{\vectorbold*{E}}^\mu_{\ \nu}}{\partial \vec{x}}  + \bar{\mathcal{L}}_{\vectorbold*{B}^\mu_{\ \nu}} \frac{\partial \bar{\vectorbold*{B}}^\mu_{\ \nu}}{\partial \vec{x}}+\cdots\;, \label{eq:con_rho_Lbar}
\end{align}
where $\cdots$ refers to higher-order terms in derivatives in case the more general action~(\ref{eq:action_uni}) involving the derivative operators~$\pounds_{\vecb{n}}$ and $\vecb{D}_\mu$ is considered. 
Note that on the RHS of the above equation, there is no spatial derivative acting on $\bar{\mathcal{L}}$ since the action is not allowed to explicitly depend on $\vec{x}$.
Additionally, applying the chain rule to the spatial derivatives of $\bar{\mathcal{L}}_{\vectorbold*{g}^{\tau\tau}}$ and $\bar{\mathcal{L}}_{\vecb{Z}^I{}^\mu_{\ \nu}}$ leads to
\begin{align}
\frac{\partial}{\partial \vec{x}} \bar{\mathcal{L}}_{\vectorbold*{g}^{\tau\tau}} &= \bar{\mathcal{L}}_{\vectorbold*{g}^{\tau\tau}\vectorbold*{g}^{\tau\tau}}\frac{\partial \bar{\vectorbold*{g}}^{\tau\tau} }{\partial \vec{x}} + \bar{\mathcal{L}}_{\vectorbold*{g}^{\tau\tau}\vectorbold*{Z}^J{}^\mu_{\ \nu}}\frac{\partial \bar{\vectorbold*{Z}}^J{}^\mu_{\ \nu}}{\partial \vec{x}}+\cdots\;, \\ 
\frac{\partial}{\partial \vec{x}} \bar{\mathcal{L}}_{\vectorbold*{Z}^I{}^\mu_{\ \nu}} &= \bar{\mathcal{L}}_{\vectorbold*{g}^{\tau\tau} \vectorbold*{Z}^I{}^\mu_{\ \nu} }\frac{\partial \bar{\vectorbold*{g}}^{\tau\tau} }{\partial \vec{x}} +  \bar{\mathcal{L}}_{\vectorbold*{Z}^I{}^\mu_{\ \nu}\vectorbold*{Z}^J{}^\alpha_{\ \beta}}\frac{\partial \bar{\vectorbold*{Z}}^J{}^\alpha_{\ \beta}}{\partial \vec{x}}+\cdots\;. \label{eq:con_rho_R}
\end{align}
Straightforwardly, one can apply the chain rule to the spatial derivatives of other higher-order terms (e.g., $\bar{\mathcal{L}}_{\vecb{g}^{\tau\tau}\vecb{g}^{\tau\tau}} $ and $\bar{\mathcal{L}}_{\vecb{g}^{\tau\tau}\vecb{Z}^I{}^{\mu}_{\ \nu}}$) to obtain their corresponding consistency relations. 
In this way, we obtain a set of infinitely many consistency relations. 
We see that these relations provide a connection among the Taylor coefficients order by order in perturbations; the coefficients cannot be arbitrarily chosen. 
We now impose this set of consistency relations to make sure that the 3d diffeomorphism invariance is preserved. 
In fact, we see that in the case where all the background quantities are independent of $\vec{x}$, e.g., cosmological background~\cite{Aoki:2021wew}, this set of relations is trivially satisfied. 
In addition, when the gauge coupling is set to zero, this kind of consistency relations recovers the one found in \cite{Mukohyama:2022enj,Mukohyama:2022skk} as expected. 

Let us discuss the second set of consistency relations. This set is obtained by applying the chain rule to the time derivative of the Taylor coefficients in (\ref{eq:Taylor_first}) and (\ref{eq:Taylor_second}). Applying the chain rule to the time derivative of $\bar{\mathcal{L}}(\tau, \vec{x})$, we obtain
\begin{align}
\frac{\partial}{\partial \tau} \bar{\mathcal{L}}(\tau,\vec{x})  = \ & \frac{\rm d}{{\rm d} \tau} \bar{\mathcal{L}}(\vectorbold*{g}^{\tau\tau}, \vecb{K}^\mu_{\ \nu},{}^{(3)}\!\vecb{R}^\mu_{\ \nu},\vectorbold*{E}^\mu_{\ \nu}, \vectorbold*{B}^\mu_{\ \nu})  \bigg|_{\rm BG} \nonumber \\
= \  &\bar{\mathcal{L}}_{\vectorbold*{g}^{\tau\tau}}\frac{\partial \bar{\vectorbold*{g}}^{\tau\tau} }{\partial \tau}
+ \bar{\mathcal{L}}_{\vectorbold*{K}^\mu_{\ \nu}}\frac{\partial \bar{\vectorbold*{K}}^\mu_{\ \nu}}{\partial \tau}   + \bar{\mathcal{L}}_{{}^{(3)}\!\vectorbold*{R}^\mu_{\ \nu}}\frac{\partial{}^{(3)}\!\bar{\vectorbold*{R}}^\mu_{\ \nu}}{\partial \tau}
+ \bar{\mathcal{L}}_{\vectorbold*{E}^\mu_{\ \nu}} \frac{\partial \bar{\vectorbold*{E}}^\mu_{\ \nu}}{\partial \tau}  + \bar{\mathcal{L}}_{\vectorbold*{B}^\mu_{\ \nu}} \frac{\partial \bar{\vectorbold*{B}}^\mu_{\ \nu}}{\partial \tau}+\cdots\;.
\label{eq:con_L_tau}
\end{align}
Notice that on the RHS of the above equation, there is no time derivative acting on $\bar{\mathcal{L}}$ since the EFT action is not allowed to explicitly depend on $\tau$.
Similarly, the chain rule applied to the time derivatives of $\bar{\mathcal{L}}_{\vectorbold*{g}^{\tau\tau}}$ and $\bar{\mathcal{L}}_{\vectorbold*{Z}^I{}^\mu_{\ \nu}}$ yields 
\begin{align}
\frac{\partial}{\partial \tau} \bar{\mathcal{L}}_{\vectorbold*{g}^{\tau\tau}} &= \bar{\mathcal{L}}_{\vectorbold*{g}^{\tau\tau}\vectorbold*{g}^{\tau\tau}}\frac{\partial \bar{\vectorbold*{g}}^{\tau\tau} }{\partial \tau} + \bar{\mathcal{L}}_{\vectorbold*{g}^{\tau\tau}\vectorbold*{Z}^J{}^\mu_{\ \nu}}\frac{\partial \bar{\vectorbold*{Z}}^J{}^\mu_{\ \nu}}{\partial \tau}+\cdots\;, \\ 
\frac{\partial}{\partial \tau} \bar{\mathcal{L}}_{\vectorbold*{Z}^I{}^\mu_{\ \nu}} &= \bar{\mathcal{L}}_{\vectorbold*{g}^{\tau\tau} \vectorbold*{Z}^I{}^\mu_{\ \nu} }\frac{\partial \bar{\vectorbold*{g}}^{\tau\tau} }{\partial \tau} +  \bar{\mathcal{L}}_{\vectorbold*{Z}^I{}^\mu_{\ \nu}\vectorbold*{Z}^J{}^\alpha_{\ \beta}}\frac{\partial \bar{\vectorbold*{Z}}^J{}^\alpha_{\ \beta}}{\partial \tau}+\cdots\;. \label{eq:con_tau_R}
\end{align}
As before, one can straightforwardly apply the chain rule to the time derivatives of other terms, for instance, $\bar{\mathcal{L}}_{\vecb{g}^{\tau\tau} \vecb{Z}^I{}^\mu_{\ \nu}}$, to obtain their corresponding consistency relations. 
In this way, one again obtains a set of infinitely many consistency relations. 
This set of consistency relations must be imposed on the Taylor coefficients in order for the EFT to be invariant under the residual $U(1)$ symmetry~\eqref{eq:combined_U(1)_time}.  
We note that in the case of the EFT of scalar-tensor theories discussed in \cite{Mukohyama:2022enj,Mukohyama:2022skk} where the gauge coupling~$g_M$ vanishes,
this set of consistency relations is trivially satisfied since the EFT action is allowed to explicitly depend on the time coordinate~$\tau$ unless the underlining theory has the shift symmetry.
Moreover, once the background is given by the FLRW metric, the relations~(\ref{eq:con_L_tau})--(\ref{eq:con_tau_R}) reduce to the ones obtained in \cite{Aoki:2021wew} as expected.

In the next Subsection, we will write down the EFT action up to second order in perturbations and we will see that the two sets of consistency relations we have obtained are translated into relations among the EFT coefficients of vector-tensor theories on an arbitrary background.
 
\subsection{EFT action}\label{sec:EFT_action}
In this Subsection, we will write down the EFT action~(\ref{eq:action_uni2}) expanded up to the second order in perturbations. We will see that this EFT action is apparently not invariant under the 3d spatial diffeomorphism and the residual $U(1)$~\eqref{eq:combined_U(1)_time} due to the fact that perturbations in (\ref{eq:pert_geo_gauge}) are defined around a general background which depends on time and spatial coordinates. 
As pointed out in the previous Subsection, the two sets of consistency relations, Eqs.~(\ref{eq:con_rho_Lbar})--(\ref{eq:con_rho_R}) and Eqs.~(\ref{eq:con_L_tau})--(\ref{eq:con_tau_R}) ensure that the EFT action is invariant under the residual symmetries. After we write down the EFT action, we will impose these relations among the EFT coefficients.

Expanding the action~\eqref{eq:action_uni2} around a general background with perturbations~(\ref{eq:pert_geo_gauge}), we consider the following general EFT action in the unitary gauge up to second order in perturbations,
 \begin{align}
 S_{\rm EFT} =  \int  \mathrm{d}^4x \sqrt{-g} \bigg[&\frac{M^2_\star}{2}f(y)\vecb{R} - \Lambda(y) - c(y) \vecb{g}^{\tau\tau}  - d(y) \vecb{K} - \alpha(y)^\mu_{\ \nu} \vecb{K}^\nu_{\ \mu} - \beta(y)^\mu_{\ \nu} {}^{(3)}\!\vecb{R}^\nu_{\ \mu} \nonumber \\   
 &- \xi_1(y)^\mu_{\ \nu} \vecb{E}^\nu_{\ \mu} - \xi_2(y)^\mu_{\ \nu} \vecb{B}^\nu_{\ \mu}  
 + \frac{1}{2}m_2^4(y) \bigg(\frac{\delta \vecb{g}^{\tau\tau}}{-\bar{\vecb{g}}^{\tau\tau}}\bigg)^2  +  \lambda_2^I(y)^\mu_{\ \nu} \bigg( \frac{ \delta \vecb{g}^{\tau\tau} }{-\bar{\vecb{g}}^{\tau\tau}}\bigg) \delta \vecb{Z}^I{}^\nu_{\ \mu} \nonumber \\ 
 & + \frac{1}{2}\lambda_3^{IJ}(y)^{\mu \alpha}_{\ \ \ \nu \beta} \delta \vecb{Z}^I{}^\nu_{\ \mu} \delta \vecb{Z}^J{}^\beta_{\ \alpha}  \bigg] \;, \label{eq:EFT}
 \end{align}
where $y \equiv \{\tau,\vec{x}\}$ and $\vecb{R}$ denotes the 4d Ricci scalar without the total derivative term, i.e.,
 \begin{align}\label{eq:Ricci_4d}
\vecb{R} \equiv {}^{(3)}\!\vecb{R} + \vecb{K}^\mu_{\ \nu} \vecb{K}^\nu_{\ \mu} + \vecb{\omega}^\mu_{\ \nu} \vecb{\omega}^\nu_{\ \mu} - \vecb{K}^2 = R + 2 \nabla_\mu (\vecb{n}^\nu \nabla_\nu \vecb{n}^\mu - \vecb{K} \vecb{n}^\mu) \;,
 \end{align}
with $R$ being the usual 4d Ricci scalar.
Note that $\alpha(y)^\mu{}_\nu$ and $\beta(y)^\mu{}_\nu$ are assumed to be traceless (i.e., $\alpha^\mu_{\ \mu}=0$ and $\beta^\mu_{\ \mu}=0$) so that the EFT coefficients can be defined uniquely.\footnote{If $\alpha(y)^\mu{}_\nu$ and $\beta(y)^\mu{}_\nu$ had the trace parts, then they would be absorbed into $d(y)$ and $f(y)$, respectively.}
The EFT coefficients in (\ref{eq:EFT}) can generally depend on both time and spatial coordinates. 
It is useful to note that, in the case of the EFT of vector-tensor theories on the FLRW background, all the EFT coefficients are functions only of time, coefficients~$\xi_1(y)^\mu_{\ \nu}$ and $\xi_2(y)^\mu_{\ \nu}$ should vanish due to the isotropy of the background, and $\alpha(y)^\mu_{\ \nu}$ and $\beta(y)^\mu_{\ \nu}$ are redundant.
As usual, the Einstein-Hilbert term and the terms linear in perturbations affect the background dynamics for both the metric and the gauge field, resulting in the tadpole cancellation conditions (see Section~\ref{sec4_BG_decoupling}).

Let us comment on the second-order terms in (\ref{eq:EFT}).
In general, the coefficient~$\lambda_2^I{}^\mu_{\ \nu}$ can be decomposed as $\lambda_2^I{}^\mu_{\ \nu} = a^I\delta^\mu_\nu + \tilde{\lambda}_2^I{}^\mu_{\ \nu}$, where $a^I$ is some function of $y$ and $\tilde{\lambda}_2^I{}^\mu_{\ \nu}$ is the traceless part of $\lambda_2^I{}^\mu_{\ \nu}$. 
This ensures that the operators such as $\delta \vecb{g}^{\tau\tau} \delta \vecb{K}$ and $\delta \vecb{g}^{\tau\tau} \delta \vecb{K}^\mu_{\ \nu}$ have been taken into account. Note that, when $\delta \vecb{Z}^I{}^\mu_{\ \nu}$ is taken to be either $\delta \vecb{E}^\mu_{\ \nu}$ or $\delta \vecb{B}^\mu_{\ \nu}$, the terms that consist of the trace of $\delta \vecb{E}^\mu_{\ \nu}$ or $\delta \vecb{B}^\mu_{\ \nu}$ automatically vanish.
As for the rank-4 tensor~$\lambda_3^{IJ}{}^{\mu\alpha}_{\ \ \ \nu\beta}$ 
, it can be generally decomposed into
\begin{align}\label{eq:lambda_3_decom}
\lambda_3{}^{\mu\alpha}_{\ \ \ \nu\beta} = c_1 \delta^\mu_{\ \nu} \delta^\alpha_{\ \beta} + c_2 \delta^\mu_{\ \beta} \delta^\alpha_{\ \nu} + \hat{c}^\alpha_{\ (\beta} \delta^\mu_{\ \nu)} + \tilde{\lambda}_3{}^{\mu\alpha}_{\ \ \ \nu\beta} \;, 
\end{align}
where we have suppressed the indices~$I$ and $J$.
Here, $c_1$ and $c_2$ are functions of $y$, and $\hat{c}^\alpha_{\ \beta}$ is a rank-2 traceless tensor.
The remaining rank-4 piece is called $\tilde{\lambda}_3{}^{\mu\alpha}_{\ \ \ \nu\beta}$.
Clearly, when the first two terms in (\ref{eq:lambda_3_decom}) are contracted with $\delta \vecb{Z}^I{}^\nu_{\ \mu} \delta \vecb{Z}^J{}^\beta_{\ \alpha}$, it gives rise to the EFT operators such as $\delta \vecb{K}^2$ and $\delta \vecb{K}^\mu_{\ \nu} \delta \vecb{K}^\nu_{\ \mu}$.
Besides, the third term on the RHS of (\ref{eq:lambda_3_decom}) once contracted with $\delta \vecb{Z}^I{}^\nu_{\ \mu} \delta \vecb{Z}^J{}^\beta_{\ \alpha}$ generates the operators, e.g., $\delta \vecb{K} \delta \vecb{K}^\mu_{\ \nu}$. 
Finally, the contraction between the tensor~$\tilde{\lambda}_3{}^{\mu\alpha}_{\ \ \ \nu\beta}$ and $\delta \vecb{Z}^I{}^\nu_{\ \mu} \delta \vecb{Z}^J{}^\beta_{\ \alpha}$ gives the operators involving four indices, unlike the ones obtained from the first three terms in (\ref{eq:lambda_3_decom}).  
In what follows, we will not discuss an explicit form of the tensor~$\tilde{\lambda}_3{}^{\mu\alpha}_{\ \ \ \nu\beta}$ since it would not be relevant to our discussion in this paper.  

Let us now proceed with the EFT action~(\ref{eq:EFT}).
For notational simplicity, let us define the following quantities:
    \begin{align}
    \bar{\mathcal{L}}_{\vecb{K}}\equiv\frac{1}{4}\delta^\mu_{\ \nu}\bar{\mathcal{L}}_{\vecb{K}^\mu_{\ \nu}}\;, \quad
    \bar{\mathcal{L}}_{{}^{(3)}\!\vecb{R}}\equiv\frac{1}{4}\delta^\mu_{\ \nu}\bar{\mathcal{L}}_{{}^{(3)}\!\vecb{R}^\mu_{\ \nu}}\;,
    \label{trace_LK_LR}
    \end{align}
which correspond to the trace of the Taylor coefficients~$\bar{\mathcal{L}}_{\vecb{K}^\mu_{\ \nu}}$ and $\bar{\mathcal{L}}_{{}^{(3)}\!\vecb{R}^\mu_{\ \nu}}$, respectively.\footnote{The coefficient of $1/4$ in Eq.~\eqref{trace_LK_LR} has been introduced so that the traceless parts of $\bar{\mathcal{L}}_{\vecb{K}^\mu_{\ \nu}}$ and $\bar{\mathcal{L}}_{{}^{(3)}\!\vecb{R}^\mu_{\ \nu}}$ are given by $\bar{\mathcal{L}}_{\vecb{K}^\mu_{\ \nu}}-\bar{\mathcal{L}}_{\vecb{K}}\delta^\nu_{\ \mu}$ and $\bar{\mathcal{L}}_{{}^{(3)}\!\vecb{R}^\mu_{\ \nu}}-\bar{\mathcal{L}}_{{}^{(3)}\!\vecb{R}}\delta^\nu_{\ \mu}$, respectively.}
Then, the Taylor coefficients in (\ref{eq:Taylor_first}) and (\ref{eq:Taylor_second}) are related to the EFT coefficients via 
 \begin{align}
 \begin{split}
 M_\star^2 f(y) &= 2 \bar{\mathcal{L}}_{{}^{(3)}\!\vecb{R}} \;, \\ 
 \Lambda(y) &=  -\bar{\mathcal{L}} + \bar{\mathcal{L}}_{\vecb{g}^{\tau\tau}}  \bar{\vecb{g}}^{\tau\tau}
 + \bar{\mathcal{L}}_{\vecb{K}^\mu_{\ \nu}} \bar{\vecb{K}}^\mu_{\ \nu} + \bar{\mathcal{L}}_{{}^{(3)}\!\vecb{R}^\mu_{\ \nu}} {}^{(3)}\!\bar{\vecb{R}}^\mu_{\ \nu}
 + \bar{\mathcal{L}}_{{}^{(3)}\!\vecb{R}} \big(\bar{\vecb{K}}^2 - \bar{\vecb{K}}^\mu_{\ \nu} \bar{\vecb{K}}^\nu_{\ \mu} 
 \big) \\ 
 & \hspace{0.5cm} + \bar{\mathcal{L}}_{\vecb{E}^\mu_{\ \nu}} \bar{\vecb{E}}^\mu_{\ \nu} + \bar{\mathcal{L}}_{\vecb{B}^\mu_{\ \nu}} \bar{\vecb{B}}^\mu_{\ \nu} \;, \\
 c(y) &= - \bar{\mathcal{L}}_{\vecb{g}^{\tau\tau}} + \frac{g_M^2}{4(-\bar{\vecb{g}}^{\tau\tau})^2} \bar{\mathcal{L}}_{{}^{(3)}\!\vecb{R}} \bar{\vecb{B}}^\mu_{\ \nu} \bar{\vecb{B}}^\nu_{\ \mu} \;, \\
d(y) &= - \bar{\mathcal{L}}_{\vecb{K}} - \frac{3}{2} \bar{\mathcal{L}}_{{}^{(3)}\!\vecb{R}} \bar{\vecb{K}} \;, \quad
 \alpha(y)^\mu_{\ \nu} = - \bar{\mathcal{L}}_{\vecb{K}^\nu_{\ \mu}} +\bar{\mathcal{L}}_{\vecb{K}}\delta^\mu_{\ \nu}+ 2 \bar{\mathcal{L}}_{{}^{(3)}\!\vecb{R}} \bigg(\bar{\vecb{K}}^\mu_{\ \nu}-\frac{\bar{\vecb{K}}}{4}\delta^\mu_{\ \nu}\bigg) \;, \\
 \beta(y)^\mu_{\ \nu} &= -\bar{\mathcal{L}}_{{}^{(3)}\!\vecb{R}^\nu_{\ \mu}}+\bar{\mathcal{L}}_{{}^{(3)}\!\vecb{R}}\delta^\mu_{\ \nu} \;, \quad
 \xi_1(y)^\mu_{\ \nu} = -\bar{\mathcal{L}}_{\vecb{E}^\nu_{\ \mu}}\;, \\
 \xi_2(y)^\mu_{\ \nu} &= -\bar{\mathcal{L}}_{\vecb{B}^\nu_{\ \mu}} + \frac{g_M^2}{2 (-\bar{\vecb{g}}^{\tau\tau}) }\bar{\mathcal{L}}_{{}^{(3)}\!\vecb{R}}\bar{\vecb{B}}^\mu_{\ \nu} \;,  \quad
  m_2^4(y) =  (- \bar{\vecb{g}}^{\tau\tau})^2 \bar{\mathcal{L}}_{\vecb{g}^{\tau\tau}\vecb{g}^{\tau\tau}} - \frac{g_M^2}{2(-\bar{\vecb{g}^{\tau\tau}})}\bar{\mathcal{L}}_{{}^{(3)}\!\vecb{R}} \bar{\vecb{B}}^\mu_{\ \nu} \bar{\vecb{B}}^\nu_{\ \mu} \;,
 \end{split} \label{eq:EFT_para1}
 \end{align}
where we recall that $\alpha(y)^\mu{}_\nu$ and $\beta(y)^\mu{}_\nu$ were assumed to be traceless.
Also, the terms with $\lambda_2^I(y)^\mu_{\ \nu}$ and $\lambda_3^{IJ}(y)^{\mu \alpha}_{\ \ \ \nu \beta}$ are given by
    \begin{align}
    \begin{split}
    \lambda_2^I(y)^\mu_{\ \nu} \bigg( \frac{ \delta \vecb{g}^{\tau\tau} }{-\bar{\vecb{g}}^{\tau\tau}}\bigg) \delta \vecb{Z}^I{}^\nu_{\ \mu}
    &=\bar{\mathcal{L}}_{\vecb{g}^{\tau\tau} \vecb{Z}^I{}^\nu_{\ \mu}}\delta \vecb{g}^{\tau\tau}\delta \vecb{Z}^I{}^\nu_{\ \mu}
    -\frac{g_M^2}{2(-\bar{\vecb{g}}^{\tau\tau})^2}\bar{\mathcal{L}}_{{}^{(3)}\!\vecb{R}}\bar{\vecb{B}}^\mu_{\ \nu}\delta \vecb{g}^{\tau\tau} \delta \vecb{B}^\nu_{\ \mu}\;, \\
    \lambda_3^{IJ}(y)^{\mu \alpha}_{\ \ \ \nu \beta} \delta \vecb{Z}^I{}^\nu_{\ \mu} \delta \vecb{Z}^J{}^\beta_{\ \alpha}&=\bar{\mathcal{L}}_{\vecb{Z}^I{}^\nu_{\ \mu} \vecb{Z}^J{}^\beta_{\ \alpha}}\delta \vecb{Z}^I{}^\nu_{\ \mu} \delta \vecb{Z}^J{}^\beta_{\ \alpha}
    +2\bar{\mathcal{L}}_{{}^{(3)}\!\vecb{R}}(\delta\vecb{K}^2-\delta\vecb{K}^\mu_{\ \nu}\delta\vecb{K}^\nu_{\ \mu}) \\
    &\qquad -\frac{g_M^2}{2(-\bar{\vecb{g}}^{\tau\tau})}\bar{\mathcal{L}}_{{}^{(3)}\!\vecb{R}}\delta\vecb{B}^\mu_{\ \nu}\delta\vecb{B}^\nu_{\ \mu}\;,
    \end{split} \label{eq:EFT_para2}
    \end{align}
where we have defined $\delta\vecb{K}\equiv\vecb{K}-\bar{\vecb{K}}$.
Here, we have used the following relation:
 \begin{align}
 \vecb{\omega}^\mu_{\ \nu} \vecb{\omega}^\nu_{\ \mu} = \frac{g_M^2}{4 (-\bar{\vecb{g}}^{\tau\tau})} \bigg[ \bar{\vecb{B}}^\mu_{\ \nu} \bar{\vecb{B}}^\nu_{\ \mu} + 2 \bar{\vecb{B}}^\mu_{\ \nu} \delta \vecb{B}^\nu_{\ \mu} + \bar{\vecb{B}}^\mu_{\ \nu} \bar{\vecb{B}}^\nu_{\ \mu} \bigg(\frac{\delta \vecb{g}^{\tau\tau}}{-\bar{\vecb{g}}^{\tau\tau}}\bigg) + 2 \bar{\vecb{B}}^\mu_{\ \nu} \delta \vecb{B}^\nu_{\ \mu} \bigg(\frac{\delta \vecb{g}^{\tau\tau}}{-\bar{\vecb{g}}^{\tau\tau}}\bigg)& \nonumber \\ + \delta \vecb{B}^\mu_{\ \nu} \delta \vecb{B}^\nu_{\ \mu} +  \bar{\vecb{B}}^\mu_{\ \nu} \bar{\vecb{B}}^\nu_{\ \mu} \bigg(\frac{\delta \vecb{g}^{\tau\tau}}{-\bar{\vecb{g}}^{\tau\tau}}\bigg)^2 + \cdots \,&\bigg] \;, \label{eq:omega_square}
 \end{align}
which holds up to second order in perturbations.
We see that in the case where $\vecb{B}^\mu_{\ \nu}$ does not acquire background values, the quantity~$ \vecb{\omega}^\mu_{\ \nu} \vecb{\omega}^\nu_{\ \mu}$ starts at second order in perturbations and does not get contributions from $\delta \vecb{g}^{\tau\tau}$. 
Indeed, this is the case for the EFT of vector-tensor theories formulated on a cosmological background~\cite{Aoki:2021wew}.

Below, we are going to express the two sets of consistency relations obtained in the previous Subsection in terms of our EFT coefficients defined in (\ref{eq:EFT}).
In doing so, we rewrite the Taylor coefficients in terms of the EFT coefficients using Eqs.~\eqref{eq:EFT_para1} and \eqref{eq:EFT_para2}.
The first set of consistency relations~(\ref{eq:con_rho_Lbar})--(\ref{eq:con_rho_R}) ensures the invariance of the EFT action under the spatial diffeomorphism.
Among them, the first two relations can be rewritten in terms of the EFT coefficients as
  \begin{align}
    \partial_i\Lambda + \bar{\vecb{g}}^{\tau\tau} \partial_i c  - \frac{M_\star^2}{2} \big(\bar{\vecb{K}}^2 - \bar{\vecb{K}}^\mu_{\ \nu} \bar{\vecb{K}}^\nu_{\ \mu} + {}^{(3)}\!\bar{\vecb{R}}\big) \partial_i f + \bar{\vecb{K}}\partial_i d + \bar{\vecb{K}}^\mu_{\ \nu} \partial_i \alpha^\nu_{\ \mu} 
    + {}^{(3)}\!\bar{\vecb{R}}^\mu_{\ \nu} \partial_i \beta^\nu_{\ \mu} & \nonumber \\
    +\bar{\vecb{E}}^\mu_{\ \nu} \partial_i \xi^\nu_1{}_{\mu} + \bar{\vecb{B}}^\mu_{\ \nu} \partial_i \xi^\nu_2{}_{\mu} - \frac{g_M^2 M_\star^2}{8 (- \bar{\vecb{g}}^{\tau\tau})} \bar{\vecb{B}}^\mu_{\ \nu} \bar{\vecb{B}}^\nu_{\ \mu}\partial_i f&\simeq 0\;, \label{eq:con_rho1}\\
(-\bar{\vecb{g}}^{\tau\tau})\partial_i c - m_2^4 \partial_i \left[\log(-\bar{\vecb{g}}^{\tau\tau})\right] + \lambda_2^J{}^\mu_{\ \nu} \partial_i \bar{\vecb{Z}}^J{}^\nu_{\ \mu} - \frac{g_M^2M_\star^2}{8(-\bar{\vecb{g}}^{\tau\tau})} \bar{\vecb{B}}^\mu_{\ \nu} \bar{\vecb{B}}^\nu_{\ \mu} \partial_i f&\simeq 0 \;. \label{eq:con_rho2}
 \end{align}
Also, Eq.~\eqref{eq:con_rho_R} yields consistency relations associated with the spatial derivative of $d$, $\alpha^\mu_{\ \nu}$, $\beta^\mu_{\ \nu}$, $\xi^\mu_1{}_{\nu}$, and $\xi^\mu_2{}_{\nu}$.
For instance,
    \begin{align}
    \delta^\mu_{\ \nu}\partial_i d+\partial_i \alpha^\mu_{\ \nu}-M_\star^2\big(\bar{\vecb{K}}^\mu_{\ \nu}-\bar{\vecb{K}}\delta^\mu_{\ \nu}\big)\partial_i f
    -\lambda_2^1{}^\mu_{\ \nu}\partial_i \left[\log(-\bar{\vecb{g}}^{\tau\tau})\right]+\lambda_3^{1J}{}^{\mu\alpha}_{\ \ \ \nu\beta}\partial_i\bar{\vecb{Z}}^J{}^\beta_{\ \alpha}\simeq 0\;,
    \label{eq:con_rho3}
    \end{align}
where $\lambda_2^1{}^\mu_{\ \nu}$ and $\lambda_3^{1J}{}^{\mu\alpha}_{\ \ \ \nu\beta}$ are the EFT coefficients associated with $\delta\vecb{g}^{\tau\tau}\delta\vecb{K}^\nu_{\ \mu}$ and $\delta\vecb{K}^\nu_{\ \mu}\delta\vecb{Z}^J{}^\beta_{\ \alpha}$, respectively.
Note that the above set of consistency relations reduces to that of the EFT of scalar-tensor theories when the gauge coupling constant vanishes.
Furthermore, the relations~(\ref{eq:con_rho1})--(\ref{eq:con_rho3}) trivially hold on the FLRW background. 

The second set of consistency relations~(\ref{eq:con_L_tau})--(\ref{eq:con_tau_R}) guarantees that the EFT action is invariant under the residual $U(1)$ symmetry~\eqref{eq:combined_U(1)_time}. 
Among them, the first two relations written in terms of the EFT coefficients read
 \begin{align}
    \dot{\Lambda} + \bar{\vecb{g}}^{\tau\tau} \dot{c} - \frac{M_\star^2}{2} \big(\bar{\vecb{K}}^2 - \bar{\vecb{K}}^\mu_{\ \nu} \bar{\vecb{K}}^\nu_{\ \mu} + {}^{(3)}\!\bar{\vecb{R}}\big) \dot{f} + \bar{\vecb{K}}\dot{d} + \bar{\vecb{K}}^\mu_{\ \nu} \dot{\alpha}^\nu_{\ \mu} 
    + {}^{(3)}\!\bar{\vecb{R}}^\mu_{\ \nu} \dot{\beta}^\nu_{\ \mu}& \nonumber \\
    +\bar{\vecb{E}}^\mu_{\ \nu} \dot{\xi}^\nu_1{}_{\mu} + \bar{\vecb{B}}^\mu_{\ \nu} \dot{\xi}^\nu_2{}_{\mu} - \frac{g_M^2 M_\star^2}{8 (- \bar{\vecb{g}}^{\tau\tau})} \bar{\vecb{B}}^\mu_{\ \nu} \bar{\vecb{B}}^\nu_{\ \mu}\dot{f}&\simeq 0\;, \label{eq:con_tau1} \\
\dot{c} (-\bar{\vecb{g}}^{\tau\tau}) - m_2^4 \left[\log(-\bar{\vecb{g}}^{\tau\tau})\right]^{\boldsymbol{\cdot}} + \lambda_2^J{}^\mu_{\ \nu} \dot{\bar{\vecb{Z}}}^J{}^\nu_{\ \mu} - \frac{g_M^2M_\star^2}{8(-\bar{\vecb{g}}^{\tau\tau})} \bar{\vecb{B}}^\mu_{\ \nu} \bar{\vecb{B}}^\nu_{\ \mu}\dot{f} &\simeq 0 \;, \label{eq:con_tau2}
 \end{align}
where a dot denotes the derivative with respect to $\tau$.
Also, Eq.~\eqref{eq:con_tau_R} yields consistency relations associated with the time derivative of $d$, $\alpha^\mu_{\ \nu}$, $\beta^\mu_{\ \nu}$, $\xi^\mu_1{}_{\nu}$, and $\xi^\mu_2{}_{\nu}$.
For instance,
    \begin{align}
    \delta^\mu_{\ \nu}\dot{d}+\dot{\alpha}^\mu_{\ \nu}-M_\star^2\big(\bar{\vecb{K}}^\mu_{\ \nu}-\bar{\vecb{K}}\delta^\mu_{\ \nu}\big)\dot{f}
    -\lambda_2^1{}^\mu_{\ \nu}\left[\log(-\bar{\vecb{g}}^{\tau\tau})\right]^{\boldsymbol{\cdot}}+\lambda_3^{1J}{}^{\mu\alpha}_{\ \ \ \nu\beta}\dot{\bar{\vecb{Z}}}^J{}^\beta_{\ \alpha}\simeq 0\;.
    \label{eq:con_tau3}
    \end{align}
When the background metric is given by the FLRW metric, these consistency relations coincide with the ones found in \cite{Aoki:2021wew}.
Moreover, when the gauge coupling vanishes, the relations~(\ref{eq:con_tau1})--(\ref{eq:con_tau3}) guarantee the residual shift symmetry of the EFT action for the corresponding shift-symmetric scalar-tensor EFT~\cite{Finelli:2018upr}. On the other hand, if we do not impose the shift symmetry for the scalar-tensor EFT, then this set of consistency relations should not be imposed since the EFT action in this case depends explicitly on $\tau$~\cite{Mukohyama:2022enj,Mukohyama:2022skk}.

So far, we have focused on the EFT action~\eqref{eq:EFT} which is based on (\ref{eq:action_uni2}) and which includes perturbations up to the second order.
For practical use, it is useful to introduce a subclass of the EFT described by the following action:\footnote{Note that here we define the coefficients~$\gamma_1$ and $\gamma_2$ in a different way compared to Eq.~(2.68) of \cite{Aoki:2021wew}. The relations between our coefficients and their coefficients are given by $\gamma_1^{\rm ours} = \gamma_1^{\rm theirs}$ and $\gamma_2^{\rm ours} = \gamma_1^{\rm theirs} + \gamma_2^{\rm theirs}$.}
\begin{align}
S_{\rm EFT} =\ & \int \mathrm{d}^4x \sqrt{-g}\bigg[
\frac{M^2_\star}{2}f(y)\vecb{R} - \Lambda(y) - c(y) \vecb{g}^{\tau\tau} - d(y) \vecb{K} \nonumber \\   
&- \alpha(y)^\mu_{\ \nu} \vecb{K}^\nu_{\ \mu} - \beta(y)^\mu_{\ \nu} {}^{(3)}\!\vecb{R}^\nu_{\ \mu} - \xi_1(y)^\mu_{\ \nu} \vecb{E}^\nu_{\ \mu} - \xi_2(y)^\mu_{\ \nu} \vecb{B}^\nu_{\ \mu} \nonumber \\
&+\frac{1}{2}m_2^4(y) \bigg(\frac{\delta \vecb{g}^{\tau\tau}}{-\bar{\vecb{g}}^{\tau\tau}}\bigg)^2 - \frac{1}{2}M_1^3(y) \bigg(\frac{\delta \vecb{g}^{\tau\tau}}{-\bar{\vecb{g}}^{\tau\tau}}\bigg) \delta \vecb{K} - \frac{1}{2}M_2^2(y) \delta \vecb{K}^2 -\frac{1}{2}M_3^2(y)\delta\vecb{K}^\mu_{\ \nu}\delta\vecb{K}^\nu_{\ \mu}  \nonumber \\  
&+ \frac{1}{2}\mu_1^2(y) \bigg( \frac{ \delta \vecb{g}^{\tau\tau} }{-\bar{\vecb{g}}^{\tau\tau}}\bigg) \delta {}^{(3)}\!\vecb{R}  + \frac{1}{2} \mu_2(y)^\nu_{\ \mu} \bigg( \frac{ \delta \vecb{g}^{\tau\tau} }{-\bar{\vecb{g}}^{\tau\tau}}\bigg) \delta \vecb{K}^\mu_{\ \nu} + \frac{1}{2}\zeta_1(y)^\mu_{\ \nu} \bigg( \frac{ \delta \vecb{g}^{\tau\tau} }{-\bar{\vecb{g}}^{\tau\tau}}\bigg) \delta \vecb{E}^\nu_{\ \mu} \nonumber \\ 
&+ \frac{1}{2}\zeta_2(y)^\mu_{\ \nu} \bigg( \frac{ \delta \vecb{g}^{\tau\tau} }{-\bar{\vecb{g}}^{\tau\tau}}\bigg)  \delta \vecb{B}^\nu_{\ \mu} + \frac{1}{4}\gamma_1(y) \delta \vecb{E}^\mu_{\ \nu} \delta \vecb{E}^\nu_{\ \mu} + \frac{1}{4}\gamma_2(y) \delta \vecb{B}^{\mu}_{\ \nu} \delta \vecb{B}^{\nu}_{\ \mu} \nonumber \\ 
& + \frac{1}{4}\gamma_3(y)(\bar{\vecb{E}}^\mu_{\ \nu}\delta \vecb{E}^\nu_{\ \mu})^2 + \frac{1}{4}\gamma_4(y)(\bar{\vecb{B}}^\mu_{\ \nu}\delta \vecb{B}^\nu_{\ \mu})^2 + \frac{1}{4}\gamma_5(y)\bar{\vecb{E}}^\mu_{\ \nu}\bar{\vecb{B}}^\alpha_{\ \beta}\delta \vecb{E}^\nu_{\ \mu} \delta \vecb{B}^\beta_{\ \alpha} \bigg] \;.
\label{eq:EFT_GP}
\end{align}
Note that the above action amounts to a specific choice of $\lambda_2^I{}^\mu_{\ \nu}$ and $\lambda_3^{IJ}{}^{\mu\alpha}_{\ \ \ \nu\beta}$ in \eqref{eq:EFT}.
Actually, we will show in Section~\ref{sec:Dictionary} that this action accommodates (a subclass of) the GP theories.
We will also see that, as expected, the consistency relations we have found in this Subsection are automatically satisfied for the GP theories on an arbitrary background.
 
 
 \subsection{St\"{u}ckelberg trick}\label{sec:stuck}

In the previous Subsection, we have formulated a consistent EFT of vector-tensor theories in the unitary gauge on a general background. 
In particular, we have found that the invariance of the EFT action under the residual symmetries imposes the consistency relations among the EFT coefficients. 
In this Subsection, we will discuss the St\"{u}ckelberg procedure of our EFT. 
Specifically, this will be useful for analyzing the dynamics of the longitudinal mode of $A_\mu$ in the decoupling limit, see Subsection~\ref{sec:decoupling_pi}. 

The St\"{u}ckelberg trick is a well-known procedure that is used to restore the full 4d diffeomorphism invariance of the EFT action. 
The idea is to introduce a Goldstone boson field~$\pi$ that, in this vector-tensor case, non-linearly realizes both the $U(1)$ and the time diffeomorphism symmetries. 
Notice that a combination of them, given by \eqref{eq:combined_U(1)_time}, is left unbroken for the EFT in the unitary gauge. 
Here, we introduce a St\"{u}ckelberg field~$\pi$ via transformations,\footnote{Alternatively, one can introduce the field $\pi$ via $\tau \rightarrow \tau + \pi$.}
\begin{align}
A_{\tau} \rightarrow A_{\tau} + g_M^{-1} \dot{\pi} \;, \quad A_i \rightarrow A_i + g_M^{-1} \partial_i \pi \;. \label{eq:Stuck_A_mu}
\end{align}
This Goldstone boson field~$\pi$ represents the longitudinal mode of the gauge field. 
Note that the transformation above does not affect the spatial coordinates~$\vec{x}$, so that the $\vec{x}$-dependent parts of our EFT remain unchanged.
Then, from (\ref{eq:g_tautau}), it is straightforward to obtain 
\begin{align}
\vecb{g}^{\tau\tau} \rightarrow \vecb{g}^{\tau\tau} + 2 \partial^\tau \pi + 2 g_M A^\alpha \partial_\alpha \pi + (\partial_\alpha \pi)^2 \;, \label{eq:gtautau_stuck}
\end{align}
where $\partial^\tau \pi\equiv g^{\tau\alpha}\partial_\alpha\pi$ and $(\partial_\alpha \pi)^2\equiv g^{\alpha\beta} \partial_\alpha \pi \partial_\beta \pi$.
Clearly, when the gauge coupling is set to zero, the formula above reduces to the known result in the literature for the scalar-tensor EFT (see, e.g., \cite{Mukohyama:2022enj,Cusin:2017mzw}). Using (\ref{eq:gtautau_stuck}), one finds that the vector~$\vecb{n}_\mu$ and the projection tensor~$\vecb{h}_{\mu\nu}$ transform, up to second order in $\pi$, as 
\begin{align}
\vecb{n}_\mu &\rightarrow \vecb{n}_\mu + \delta_1 \vecb{n}_\mu + \delta_2 \vecb{n}_\mu \;, \label{eq:n_mu_stu} \\ 
\vecb{h}_{\mu\nu} &\rightarrow \vecb{h}_{\mu\nu} + \delta_1 \vecb{h}_{\mu\nu} + \delta_2 \vecb{h}_{\mu\nu} \;, \label{eq:h_munu_stu}
\end{align}
where we have defined 
\begin{align}
\begin{split}
\delta_1 \vecb{n}_\mu &\equiv -Q\vecb{n}_\mu - \frac{\partial_\mu \pi}{\sqrt{-\vecb{g}^{\tau\tau}}}\;, \quad
\delta_2\vecb{n}_\mu\equiv \bigg[-\frac{(\partial_\alpha\pi)^2}{2\vecb{g}^{\tau\tau}}+\frac{3}{2}Q^2\bigg]\vecb{n}_\mu+\frac{Q}{\sqrt{-\vecb{g}^{\tau\tau}}}\partial_\mu\pi\;, \\
\delta_1 \vecb{h}_{\mu\nu} &\equiv 2 \vecb{n}_{(\mu} \delta_1 \vecb{n}_{\nu)} \;, \quad 
\delta_2 \vecb{h}_{\mu\nu} \equiv \delta_1 \vecb{n}_\mu \delta_1 \vecb{n}_\nu + 2 \vecb{n}_{(\mu} \delta_2 \vecb{n}_{\nu)} \;, 
\end{split}
\end{align}
and $Q\equiv (\partial^\tau \pi + g_M A^\alpha \partial_\alpha \pi)/\vecb{g}^{\tau\tau}$.
It is also useful to find how the object~$\vecb{h}^\alpha_{\ \mu} \nabla_{\alpha} \vecb{n}_\nu$ transforms under the St\"{u}ckelberg trick. Up to second order in $\pi$, we have 
\begin{align}
\vecb{h}^\alpha_{\ \mu} \nabla_{\alpha} \vecb{n}_\nu \rightarrow 
&\vecb{h}^\alpha_{\ \mu} \nabla_{\alpha} \vecb{n}_\nu + \delta_1 \vecb{h}^\alpha_{\ \mu} \nabla_{\alpha} \vecb{n}_\nu 
+ \delta_2 \vecb{h}^\alpha_{\ \mu} \nabla_{\alpha} \vecb{n}_\nu + \vecb{h}^\alpha_{\ \mu} \nabla_{\alpha} \delta_1 \vecb{n}_\nu 
\nonumber \\ &+ \vecb{h}^\alpha_{\ \mu} \nabla_{\alpha} \delta_2 \vecb{n}_\nu + \delta_1 \vecb{h}^\alpha_{\ \mu} \nabla_{\alpha} \delta_1 \vecb{n}_\nu \;,
\end{align}
where $\delta_1\vecb{h}^\mu_{\ \nu}\equiv g^{\mu\alpha}\delta_1\vecb{h}_{\alpha\nu}$ and $\delta_2\vecb{h}^\mu_{\ \nu}\equiv g^{\mu\alpha}\delta_2\vecb{h}_{\alpha\nu}$.
We see that symmetrizing or anti-symmetrizing the object above w.r.t.~the indices~$\mu$ and $\nu$ straightforwardly gives rise to the St\"{u}ckelberg transformations for~$\vecb{K}_{\mu\nu}$ and $\vecb{\omega}_{\mu\nu}$, respectively [see Eqs.~(\ref{eq:extrinsic}) and (\ref{eq:omega})]. 
In addition, using the formulas~(\ref{eq:n_mu_stu}) and (\ref{eq:h_munu_stu}), one can derive the St\"{u}ckelberg transformations for the acceleration vector~$\vecb{a}_{\mu}$ and the orthogonal spatial Ricci tensor~${}^{(3)}\!\vecb{R}_{\mu\nu}$. 
Finally, for completeness, we express the St\"{u}ckelberg transformations for $\vecb{E}_{\mu\nu}$ and $\vecb{B}_{\mu\nu}$. 
From Eqs.~(\ref{eq:B_munu}) and (\ref{eq:E_munu}), up to second order in $\pi$, we obtain 
    \begin{align}
    \vecb{E}_{\mu\nu}&\to \vecb{E}_{\mu\nu}+\epsilon_{\mu\nu}^{\ \ \ \rho\alpha}\big(\vecb{n}_\rho\delta_1\vecb{n}^\beta+\vecb{n}^\beta\delta_1\vecb{n}_\rho+\vecb{n}_\rho\delta_2\vecb{n}^\beta+\vecb{n}^\beta\delta_2\vecb{n}_\rho+\delta_1\vecb{n}_\rho\delta_1\vecb{n}^\beta\big)F_{\alpha\beta}\;, \\
    \vecb{B}_{\mu\nu}&\to \vecb{B}_{\mu\nu}+\big(2\vecb{h}^\alpha_{\ [\mu}\delta_1\vecb{h}^\beta_{\ \nu]}+\vecb{h}^\alpha_{\ [\mu}\delta_2\vecb{h}^\beta_{\ \nu]}+2\delta_1\vecb{h}^\alpha_{\ \mu}\delta_1\vecb{h}^\beta_{\ \nu}\big)F_{\alpha\beta}\;,
    \end{align}
where $\delta_1\vecb{n}^\mu\equiv g^{\mu\alpha}\delta_1\vecb{n}_\alpha$ and $\delta_2\vecb{n}^\mu\equiv g^{\mu\alpha}\delta_2\vecb{n}_\alpha$.
Note that, by construction, the field strength tensor~$F_{\mu\nu}$ remains invariant under the transformation~(\ref{eq:Stuck_A_mu}). 
We emphasize that the results in this Subsection can be applied to any background configurations for the metric and the gauge field. 


\section{Dictionary}\label{sec:Dictionary}
In the previous Section, we have written down the unitary-gauge EFT action of perturbations on an arbitrary background in the context of vector-tensor theories and we have derived the two sets of consistency relations keeping the EFT action invariant under the residual symmetries. 
Here, we provide a dictionary between our EFT~\eqref{eq:EFT_GP}
and those of the GP theories~\cite{Heisenberg:2014rta}. 
Let us consider a subclass of the GP theory described by the following action:
 \begin{align}
 S_{\rm GP} = \int \mathrm{d}^4x \sqrt{-g}\big\{G_2(X, F, Y) + G_3(X) \nabla_\mu A^\mu + G_4(X) R + G_{4X}\left[(\nabla_\mu A^\mu)^2 - \nabla_\mu A_\nu \nabla^\nu A^\mu \right] \big\} \;, \label{eq:GP_action}
 \end{align}
where the subscript~$X$ denotes the derivative with respect to $X$, and the variables~$X$, $F$, and $Y$ are defined as
\begin{align}
X \equiv -\frac{1}{2} A_\mu A^\mu \;, \quad F \equiv -\frac{1}{4} F_{\mu\nu} F^{\mu\nu} \;, \quad Y \equiv A^\mu A^\nu F_{\mu \alpha} F_\nu^{\ \alpha} \;.
\end{align}
Note that $G_2$ is an arbitrary function of $X$, $F$, and $Y$, while $G_3$ and $G_4$ are free functions of $X$. 

It is clear that the action~(\ref{eq:GP_action}) is not invariant under $U(1)$ gauge transformation; in other words, the theory is constructed under the gauge~$\tilde{\tau}=0$. As explained in Section~\ref{sec:symmetry}, it is convenient to move to the other gauge when $A_{\mu}$ has a timelike background. The $U(1)$ symmetry can be restored by introducing a St\"{u}ckelberg field~$\tilde{\tau}$: $A_\mu \rightarrow A_\mu + g_M^{-1} \partial_\mu\tilde{\tau}$.\footnote{This is somewhat different from the previous Section in which we started from the EFT with the residual symmetries [i.e., the 3d spatial diffeomorphism and the residual $U(1)$ symmetry~\eqref{eq:combined_U(1)_time}] and restored the full symmetries by introducing the St\"{u}ckelberg field~$\pi$. Instead, here we start from a covariant theory which respects the 4d diffeomorphism invariance, but not the $U(1)$ symmetry.}
We now consider the unitary gauge where the time coordinate~$\tau$ is chosen to coincide with $\tilde{\tau}$, so that one obtains 
\begin{align}
A_\mu \rightarrow g_M^{-1} \vecb{\delta}^\tau_{\ \mu} = -\frac{\sqrt{-\vecb{g}^{\tau\tau}}}{g_M} \vecb{n}_\mu \;,
\end{align}
where the arrow here refers to a mapping to the unitary-gauge quantity. 
Note that now the residual symmetries of GP theories are the 3d spatial diffeomorphism and the residual $U(1)$ symmetry~\eqref{eq:combined_U(1)_time}, which are the same symmetries of our EFT constructed in the unitary gauge. 
Additionally, the variables~$X$, $F$, and $Y$ in the unitary gauge are given by
\begin{align}
X = - \frac{\vecb{g}^{\tau\tau}}{2 g_M^2} \;, \quad F = -\frac{1}{4} (\vecb{E}^\mu_{\ \nu} \vecb{E}^\nu_{\ \mu} - \vecb{B}^\mu_{\ \nu} \vecb{B}^\nu_{\ \mu}) \;, \quad Y = \frac{\vecb{g}^{\tau\tau}}{2 g_M^2} \vecb{E}^\mu_{\ \nu} \vecb{E}^\nu_{\ \mu} \;.
\end{align}
Performing integration by parts, the action~(\ref{eq:GP_action}) can be written in terms of our building blocks as 
\begin{align}
S_{\rm GP} = \int \mathrm{d}^4x \sqrt{-g} \big[G_2 + F_3 \vecb{K} + G_4 {}^{(3)}\!\vecb{R} + (G_4 - 2X G_{4X}) (\vecb{K}^\mu_{\ \nu}\vecb{K}^\nu_{\ \mu} - \vecb{K}^2 + \vecb{\omega}^\mu_{\ \nu} \vecb{\omega}^\nu_{\ \mu})\big] \;, \label{eq:GP_Geo}
\end{align}
where we have defined $F_3$ such that the relation~$F_{3X} = -(2 X)^{1/2} G_{3X}$ holds. 
Straightforwardly, expanding the action~(\ref{eq:GP_Geo}) around an arbitrary background up to second order in perturbations [defined in Eq.~(\ref{eq:pert_geo_gauge})], we can read off the 
EFT coefficients. For example, one Taylor expands the function~$G_2(X,F,Y)$ as 
\begin{align}
G_2(X,F,Y) = \ & \bar{G}_2 -  (\bar{G}_{2X} - \bar{G}_{2Y} \bar{\vecb{E}}^\mu_{\ \nu}\bar{\vecb{E}}^\nu_{\ \mu}) \frac{\delta \vecb{g}^{\tau\tau}}{2 g_M^2} - \frac{1}{2}  (4 \bar{X} \bar{G}_{2Y} + \bar{G}_{2F}) \bar{\vecb{E}}^\mu_{\ \nu} \delta \vecb{E}^\nu_{\ \mu} + \frac{1}{2} \bar{G}_{2F} \bar{\vecb B}^\mu_{\ \nu}\delta \vecb{B}^\nu_{\ \mu} \nonumber \\ 
& + [\bar{G}_{2XX} - 2 \bar{G}_{2XY} \bar{\vecb{E}}^\mu_{\ \nu}\bar{\vecb{E}}^\nu_{\ \mu} + \bar{G}_{2YY} (\bar{\vecb{E}}^\mu_{\ \nu} \bar{\vecb{E}}^\nu_{\ \mu})^2] \frac{(\delta \vecb{g}^{\tau\tau})^2}{8 g_M^4} - \frac{1}{4} ( \bar{G}_{2F} + 4 \bar{X} \bar{G}_{2Y}) \delta \vecb{E}^\mu_{\ \nu} \delta \vecb{E}^\nu_{\ \mu} \nonumber \\
&+ \frac{1}{4} \bar{G}_{2F} \delta \vecb{B}^\mu_{\ \nu} \delta \vecb{B}^\nu_{\ \mu} + \frac{1}{4g_M^2} (\bar{G}_{2FY}\bar{\vecb{E}}^\alpha_{\ \beta}\bar{\vecb{E}}^\beta_{\ \alpha} - \bar{G}_{2XF} )\bar{\vecb{B}}^\mu_{\ \nu}  \delta \vecb{g}^{\tau\tau} \delta \vecb{B}^\nu_{\ \mu} \nonumber \\ 
&+ \frac{1}{4g_M^2}(4\bar{G}_{2Y} - 4 \bar{X} \bar{G}_{2YY} \bar{\vecb{E}}^\alpha_{\ \beta}\bar{\vecb{E}}^\beta_{\ \alpha} -  \bar{G}_{2FY} \bar{\vecb{E}}^\alpha_{\ \beta}\bar{\vecb{E}}^\beta_{\ \alpha} + 4 \bar{X} \bar{G}_{2XY} + \bar{G}_{2XF}) \bar{\vecb{E}}^\mu_{\ \nu} \delta \vecb{g}^{\tau\tau} \delta \vecb{E}^\nu_{\ \mu} \nonumber \\ 
& + \frac{1}{8} (16 \bar{X}^2 \bar{G}_{2 YY} +  \bar{G}_{2FF} + \bar{X} \bar{G}_{2FY}) (\bar{\vecb{E}}^\mu_{\ \nu}\delta \vecb{E}^\nu_{\ \mu})^2  \nonumber \\ 
&+ \frac{1}{8} \bar{G}_{2FF} (\bar{\vecb{B}}^\mu_{\ \nu} \delta \vecb{B}^\nu_{\ \mu} )^2
- \frac{1}{4}  (4 \bar{X} \bar{G}_{2FY} + \bar{G}_{2FF}) \bar{\vecb{E}}^\mu_{\ \nu} \bar{\vecb{B}}^\alpha_{\ \beta} \delta \vecb{E}^\nu_{\ \mu} \delta \vecb{B}^\beta_{\ \alpha} + \cdots \;. \label{eq:G2_dict}
\end{align}
Similarly, one can perform the Taylor expansion for the other terms in (\ref{eq:GP_Geo}) up to the second order. We have
\begin{align}
F_3 \vecb{K} &= \bar{F}_3 \bar{\vecb K} -\bar{\vecb{K}} \bar{F}_{3X}  \frac{\delta \vecb{g}^{\tau\tau}}{2 g_M^2}   + \bar{F}_3 \delta \vecb{K} + \bar{\vecb{K}} \bar{F}_{3XX} \frac{(\delta \vecb{g}^{\tau\tau})^2}{8 g_M^4} - \bar{F}_{3X} \frac{\delta \vecb{g}^{\tau\tau}}{2 g_M^2} \delta \vecb{K}  \;, \label{eq:F3_dict} \\ 
G_4 {}^{(3)}\!\vecb{R} &= \bar{G}_4 {}^{(3)}\!\bar{\vecb R} - {}^{(3)}\!\bar{\vecb{R}}\,\bar{G}_{4X}  \frac{\delta \vecb{g}^{\tau\tau}}{2 g_M^2}   + \bar{G}_4 \delta {}^{(3)}\!\vecb{R} + {}^{(3)}\!\bar{\vecb{R}} \,\bar{G}_{4XX} \frac{(\delta \vecb{g}^{\tau\tau})^2}{8 g_M^4} - \bar{G}_{4X} \frac{\delta \vecb{g}^{\tau\tau}}{2 g_M^2} \delta {}^{(3)}\!\vecb{R} \;, \label{eq:G4_dict}
\end{align}
and the expansion of the last term in the RHS of (\ref{eq:GP_Geo}) gives 
\begin{align}
\vecb{K}^\mu_{\ \nu}\vecb{K}^\nu_{\ \mu} - \vecb{K}^2 + \vecb{\omega}^\mu_{\ \nu} \vecb{\omega}^\nu_{\ \mu} = \ & \bar{\vecb{K}}^\mu_{\ \nu}\bar{\vecb{K}}^\nu_{\ \mu} - \bar{\vecb{K}}^2 + \bar{\vecb{\omega}}^\mu_{\ \nu} \bar{\vecb{\omega}}^\nu_{\ \mu} + 2 \bar{\vecb K}^\mu_{\ \nu} \delta \vecb{K}^\nu_{\ \mu} - 2 \bar{\vecb K} \delta \vecb{K} + \frac{g_M^2}{2(-\bar{\vecb g}^{\tau\tau})} \bar{\vecb B}^\mu_{\ \nu} \delta \vecb{B}^\nu_{\ \mu} \nonumber \\
&+ \frac{g_M^2}{4(-\bar{\vecb g}^{\tau\tau})^2} \bar{\vecb{B}}^\mu_{\ \nu}\bar{\vecb{B}}^\nu_{\ \mu} \delta \vecb{g}^{\tau\tau} + \delta \vecb{K}^\mu_{\ \nu}\delta \vecb{K}^\nu_{\ \mu} - \delta {\vecb K}^2 + \frac{g_M^2}{2(-\bar{\vecb g}^{\tau\tau})^2} \bar{\vecb B}^\mu_{\ \nu} \delta \vecb{g}^{\tau\tau} \delta \vecb{B}^\nu_{\ \mu}  \nonumber \\
&+ \frac{g_M^2}{4(-\bar{\vecb g}^{\tau\tau})} \delta \vecb{B}^\mu_{\ \nu} \delta \vecb{B}^\nu_{\ \mu} + \frac{g_M^2}{4(-\bar{\vecb g}^{\tau\tau})^3} \bar{\vecb{B}}^\mu_{\ \nu} \bar{\vecb{B}}^\nu_{\ \mu} (\delta \vecb{g}^{\tau\tau})^2 \;.
\end{align}
Moreover, the expansion of the function~$G_4 - 2X G_{4X}$ is similar to the one of $F_3(X)$ or $G_4(X)$ [Eqs.~(\ref{eq:F3_dict}) and (\ref{eq:G4_dict})], which gives contributions to the terms that are zeroth-, first-, and second-order in $\delta \vecb{g}^{\tau\tau}$. Notice that, apart from the expansion of $G_2$ [Eq.~(\ref{eq:G2_dict})], the terms~$\delta \vecb{E}^\nu_{\ \mu} \delta \vecb{E}^\beta_{\ \alpha}$, $\delta \vecb{B}^\nu_{\ \mu} \delta \vecb{B}^\beta_{\ \alpha}$, and $\delta \vecb{E}^\nu_{\ \mu} \delta \vecb{B}^\beta_{\ \alpha}$ are not present in the expansion of other terms in (\ref{eq:GP_Geo}).

With the above Taylor expansions, the dictionary between the EFT coefficients in \eqref{eq:EFT_GP} and those of GP theories is given by  
\begin{align}
\begin{split}
M_\star^2 f &= 2 \bar{G}_4 \;, \\  
\Lambda &= -\bar{G}_2 + \bar{X} \bar{G}_{2X} + \bar{X} \bar{\vecb{K}} \bar{F}_{3X} - \bar{X}(\bar{\vecb{K}}^\mu_{\ \nu} \bar{\vecb{K}}^\nu_{\ \mu} - \bar{\vecb{K}}^2) (3 \bar{G}_{4X} + 2 \bar{X} \bar{G}_{4XX}) + \bar{X} {}^{(3)}\!\bar{\vecb{R}}\, \bar{G}_{4X} \\ 
& \hspace{5mm} + \frac{1}{8} \bar{\vecb{B}}^\mu_{\ \nu} \bar{\vecb{B}}^\nu_{\ \mu} (4 \bar{G}_{2F} - \bar{G}_{4X} - 2 \bar{X} \bar{G}_{4X}) - \frac{1}{2} \bar{\vecb{E}}^\mu_{\ \nu} \bar{\vecb{E}}^\nu_{\ \mu} (\bar{G}_{2F} + 6 \bar{X} \bar{G}_{2Y}) \;, \\ 
c &= \frac{\bar{X}}{(-\bar{\vecb{g}}^{\tau\tau})} \big[\bar{G}_{2X} + \bar{\vecb{K}} \bar{F}_{3X} + {}^{(3)}\!\bar{\vecb{R}}\, \bar{G}_{4X} - (\bar{\vecb{K}}^\mu_{\ \nu} \bar{\vecb{K}}^\nu_{\ \mu} - \bar{\vecb{K}}^2) (\bar{G}_{4X} + 2 \bar{X} \bar{G}_{4XX})  \big] \\ 
&\hspace{5mm}+ \frac{\bar{\vecb{B}}^\mu_{\ \nu} \bar{\vecb{B}}^\nu_{\ \mu} }{8(- \bar{\vecb{g}}^{\tau\tau})} (\bar{G}_{4X} - 2 \bar{X} \bar{G}_{4XX}) - \frac{\bar{X}}{(- \bar{\vecb{g}}^{\tau\tau})} \bar{G}_{2Y} \bar{\vecb{E}}^\mu_{\ \nu} \bar{\vecb{E}}^\nu_{\ \mu} \;, \\
d &= -\bar{F}_3 - 3 \bar{X} \bar{\vecb{K}} \bar{G}_{4X} \;, \quad \alpha^\mu_{\ \nu} = 4 \bar{X} \bar{G}_{4X} \bigg(\bar{\vecb{K}}^\mu_{\ \nu}-\frac{\bar{\vecb{K}}}{4}\delta^\mu_{\ \nu}\bigg) \;, \\
\xi^\mu_{1}{}_{\nu} &= \frac{1}{2} \bar{\vecb E}^\mu_{\ \nu} (\bar{G}_{2F} + 4 \bar{X} \bar{G}_{2Y}) \;, \quad
\xi^\mu_{2}{}_{\nu} = -\frac{1}{2}\bar{\vecb{B}}^\mu_{\ \nu} (\bar{G}_{2F}-\bar{G}_{4X})\;, \\ 
m_2^4 &= \bar{X}^2 \bar{G}_{2XX} + \bar{X}^2 \bar{\vecb{K}} \bar{F}_{3XX} + {}^{(3)}\!\bar{\vecb{R}}\,\bar{X}^2 \bar{G}_{4XX} - \bar{X}^2 (\bar{\vecb{K}}^\mu_{\ \nu}\bar{\vecb{K}}^\nu_{\ \mu} - \bar{\vecb{K}}^2) (3 \bar{G}_{4XX} + 2 \bar{X} \bar{G}_{4XXX}) \\ 
&\hspace{5mm}- \frac{1}{8} \bar{\vecb{B}}^\mu_{\ \nu} \bar{\vecb{B}}^\nu_{\ \mu} (2 \bar{G}_{4X} - \bar{X} \bar{G}_{4XX} + 2 \bar{X}^2 \bar{G}_{4XXX}) - \bar{X}^2 \bar{\vecb{E}}^\mu_{\ \nu} \bar{\vecb{E}}^\nu_{\ \mu}  (2 \bar{G}_{2XY} - \bar{\vecb{E}}^\alpha_{\ \beta} \bar{\vecb{E}}^\beta_{\ \alpha} \bar{G}_{2YY}) \;, \\
M_1^3 &= 2 \bar{X} \big[\bar{F}_{3X}  + 2\bar{\vecb{K}} (\bar{G}_{4X} + 2 \bar{X} \bar{G}_{4XX}) \big] \;, \quad M_2^2 = -4 \bar{X} \bar{G}_{4X} \;, \quad M_3^2 = 4 \bar{X} \bar{G}_{4X} \;, \\
\mu_1^2 &= - 2 \bar{X}  \bar{G}_{4X} \;, \quad
\mu^\mu_{2}{}_{\nu}  = 4 \bar{X} (\bar{G}_{4X} + 2 \bar{X} \bar{G}_{4XX}) \bar{\vecb{K}}^\mu_{\ \nu} \;, \\ 
 \zeta^\mu_{1}{}_{\nu}  &= \bar{X} \bar{\vecb{E}}^\mu_{\ \nu} \big[4 \bar{G}_{2Y} + \bar{G}_{2XF} + 4 \bar{X} \bar{G}_{2XY} - \bar{\vecb{E}}^\alpha_{\ \beta} \bar{\vecb{E}}^\beta_{\ \alpha} (\bar{G}_{2FY} + 4 \bar{X} \bar{G}_{2 YY}) \big] \;,  \\
\zeta^\mu_{2}{}_{\nu} &= -\frac{1}{2} \bar{\vecb{B}}^\mu_{\ \nu}\big[\bar{G}_{4X} - 2 \bar{X} \bar{G}_{4XX} + 2 \bar{X} \bar{G}_{2XF} - 2 \bar{X} \bar{G}_{2FY} \bar{\vecb{E}}^\alpha_{\ \beta} \bar{\vecb{E}}^\beta_{\ \alpha}  \big] \;, \\
\gamma_1 &= -\bar{G}_{2F} - 4 \bar{X}\bar{G}_{2Y}\;, \quad
\gamma_2 = - \bar{G}_{4X} + \bar{G}_{2F}\;, \quad
\gamma_3 = \frac{1}{2}(16 \bar{X}^2 \bar{G}_{2 YY} +  \bar{G}_{2FF} + \bar{X} \bar{G}_{2FY})\;, \\ 
\gamma_4 &= \frac{1}{2}\bar{G}_{2FF}\;, \quad 
\gamma_5 = - 4 \bar{X} \bar{G}_{2FY} - \bar{G}_{2FF} \;.
\end{split} \label{eq:dictionary}
\end{align}
The parameter~$\beta^\mu_{\ \nu}$ (i.e., the coefficient of the tadpole term~${}^{(3)}\!\vecb{R}^\nu_{\ \mu}$) is vanishing in this subclass of the GP theories.
Note that, when the backgrounds of $\vecb{E}^\mu_{\ \nu}$ and $\vecb{B}^\mu_{\ \nu}$ are absent, the dictionary above reduces to the one found in \cite{Aoki:2021wew}.
Moreover, one can straightforwardly check that the dictionary~(\ref{eq:dictionary}) automatically satisfies the two sets of consistency relations [Eqs.~(\ref{eq:con_rho1})--(\ref{eq:con_rho3}) and Eqs.~(\ref{eq:con_tau1})--(\ref{eq:con_tau3})] obtained in Subsection~\ref{sec:EFT_action}.

\section{Static and spherically symmetric background}\label{sec4_BG_decoupling}
In this Section, we apply our general vector-tensor EFT, which is applicable to any background as far as the vector field is non-vanishing and timelike, to the particular case of static and spherically symmetric background.

\subsection{Background dynamics}\label{sec:gravity}

We start with the static and spherically symmetric background metric in Schwarzschild coordinates,
	\begin{align}\label{eq:bg_metric_rt}
		{\rm d}s^2 = -A(r) {\rm d}t^2 + \frac{{\rm d}r^2}{B(r)} + r^2 \gamma_{ab}{\rm d}x^a{\rm d}x^b \;,
	\end{align}
where $A(r)$ and $B(r)$ are functions of the areal radius~$r$, and $\gamma_{ab}$ represents the metric on a unit two-sphere (i.e., $\gamma_{ab}{\rm d}x^a{\rm d}x^b = {\rm d}\theta^2 + \sin^2\theta {\rm d}\phi^2$), with $a,b=\{\theta,\phi\}$. 
Notice that here we do not assume the condition~$A(r) = B(r)$ from the outset, but we will explicitly specify when such a condition is assumed. As we will see, it is more convenient to work with the so-called Lema\^{\i}tre coordinates~\cite{Lemaitre:1933gd,Mukohyama:2005rw,Khoury:2020aya,Takahashi:2021bml}, in which the background metric~\eqref{eq:bg_metric_rt} takes the form
\begin{align}\label{eq:metric_bg}
{\rm d}s^2 = -{\rm d}\tau^2 + [1 - A(r)] {\rm d}\rho^2+ r^2 \gamma_{ab}{\rm d}x^a{\rm d}x^b \;,
\end{align}
through the coordinate transformations
\begin{align}\label{eq:coord_tranf}
{\rm d}\tau = {\rm d}t + \sqrt{\frac{1 - A}{AB}}~{\rm d}r \;, \qquad 
{\rm d}\rho = {\rm d}t + \frac{{\rm d}r}{\sqrt{AB(1 - A)}} \;.
\end{align}
Applying the above transformations, we arrive at the conclusion that the areal radius~$r$ is a function of $\rho-\tau$ and
\begin{align}
\partial_\rho r = -\dot{r} = \sqrt{\frac{B(1 - A)}{A}} \;. \label{eqn:derivatives_r}
\end{align}
Clearly, the expression above indicates the fact that the $\rho$- and $\tau$-derivatives of $r$ are functions of $r$.

For the sake of simplicity, we work in the Einstein frame where $f=1$, which can be realized by means of a conformal transformation of the metric. Additionally, for simplicity
we assume that $\alpha(y)^\mu_{\ \nu} = \alpha(y) (\bar{\vecb K}^\mu_{\ \nu}-\bar{\vecb{K}}\delta^\mu_{\ \nu}/4)$, $\xi_1(y)^\mu_{\ \nu}{} = \xi_1(y) \bar{\vecb E}^\mu_{\ \nu}$, $\xi_2(y)^\mu_{\ \nu} = \xi_2(y) \bar{\vecb B}^\mu_{\ \nu}$, and $\beta(y)^\mu_{\ \nu}=0$.\footnote{If a covariant theory contains operators such as $F_{\mu\nu} \tilde{F}^{\mu\nu}$ and $(F_{\mu\nu} \tilde{F}^{\mu\nu})^2$ with $\tilde{F}_{\mu\nu}$ being the 4d dual of the field strength tensor~$F_{\mu\nu}$, the coefficients~$\xi_1^\mu{}_{\nu}$ and $\xi_2^\mu{}_{\nu}$ can be proportional to the backgrounds of $\vecb{B}^\mu_{\ \nu}$ and $\vecb{E}^\mu_{\ \nu}$, respectively. Moreover, if a concrete theory contains a higher-derivative operator of the form~$\vecb{K}^\mu_{\ \nu} {}^{(3)}\!\vecb{R}^\nu_{\ \mu}$, then $\alpha^\mu_{\ \nu}$ acquires a term that is proportional to the traceless part of ${}^{(3)}\!\bar{\vecb{R}}^\mu_{\ \nu}$, and one also has $\beta^\mu_{\ \nu}\propto \bar{\vecb{K}}^\mu_{\ \nu}$.}
Note that all of these assumptions can be realized in the subclass of the GP theories studied in Section~\ref{sec:Dictionary}, see Eq.~(\ref{eq:dictionary}).
Finally, the staticity and the spherical symmetry of the background implies that the EFT coefficients are functions of the areal radius~$r = r(\rho - \tau)$. 
With these assumptions, the background dynamics is described by the tadpole action that only includes the terms which contribute to the first order in perturbations in the action~\eqref{eq:EFT_GP}:
 \begin{align}
 S_{{\rm EFT}}^{\rm tadpole} =  \int  \mathrm{d}^4x \sqrt{-g} \bigg[\frac{M^2_\star}{2}\vecb{R} - \Lambda(r) - c(r) \vecb{g}^{\tau\tau}  &- \tilde{d}(r) \vecb{K} - \alpha(r) \bar{\vecb K}^\mu_{\ \nu} \vecb{K}^\nu_{\ \mu}  \nonumber \\
 &- \xi_1(r) \bar{\vecb E}^\mu_{\ \nu} \vecb{E}^\nu_{\ \mu} - \xi_2(r) \bar{\vecb B}^\mu_{\ \nu} \vecb{B}^\nu_{\ \mu}   \bigg] \;, \label{eq:tad_EFT}
 \end{align}
with $\tilde{d}(r)\equiv d(r)-\alpha(r)\bar{\vecb{K}}/4$.

Then, let us write down the infinitesimal variations of some geometrical quantities relevant to derive the background equations of motion below. Under the infinitesimal transformation~$g^{\mu\nu} \rightarrow g^{\mu\nu} + \delta g^{\mu\nu}$, we find
 \begin{align}
 \delta \vecb{g}^{\tau\tau} &= \vecb{\delta}^\tau_\mu \vecb{\delta}^\tau_\nu \delta g^{\mu \nu} \;, \quad 
 \delta \vecb{n}_\mu = \frac{1}{2} \vecb{n}_\mu \vecb{n}_\alpha \vecb{n}_{\beta} \delta g^{\alpha \beta} \;, \quad
 \delta \vecb{h}_{\mu\nu} = (\vecb{n}_\mu \vecb{n}_\nu \vecb{n}_\alpha \vecb{n}_\beta - g_{\mu \alpha} g_{\nu \beta} ) \delta g^{\alpha \beta} \;,
 \end{align}
where we have used the formulas~(\ref{eq:g_tautau})--(\ref{eq:proj_hmunu}).
Using the above expressions, it is straightforward to obtain the variations of the terms~$\vecb{K}$, $\vecb{K}_{\mu\nu}$, $\vecb{E}_{\mu\nu}$, and $\vecb{B}_{\mu\nu}$ with respect to the metric. Therefore, the Einstein equation (i.e., the tadpole cancellation condition of the metric) derived from the action~(\ref{eq:tad_EFT}) is given by
\begin{align}\label{eq:eins}
M^2_\star \bar{G}_{\mu\nu}  = \bar{T}_{\mu\nu}  \;,
\end{align}
where $\bar{G}_{\mu\nu}$ is the background Einstein tensor whose non-vanishing components read 
\begin{equation}\label{Gmunu-components}
		\begin{split}
			\bar{G}^\tau{}_\tau&=-\frac{[r(1-B)]'}{r^2}+\frac{1-A}{r}\left(\frac{B}{A}\right)'\;, \\
			\bar{G}^\tau{}_\rho&=-\frac{1-A}{r}\left(\frac{B}{A}\right)'\;, \\
			\bar{G}^\rho{}_\rho&=-\frac{[r(1-B)]'}{r^2}-\frac{1}{r}\left(\frac{B}{A}\right)'\;, \\
			\bar{G}^\theta{}_\theta&=\frac{B(r^2A')'}{2r^2A}+\frac{(r^2A)'}{4r^2}\left(\frac{B}{A}\right)'\;,
		\end{split}
	\end{equation}
and the background stress-energy tensor~$\bar{T}_{\mu\nu}$ associated with the tadpole terms in (\ref{eq:tad_EFT}) reads\footnote{It is interesting to note that in comparison with the EFT of scalar-tensor theories on black hole background~\cite{Mukohyama:2022enj,Mukohyama:2022skk}, apart from the extra contributions coming from the operators~$\xi_1$ and $\xi_2$, one can simply obtain the stress-energy tensor of the present case by replacing the 3d spacelike quantities such as $n_{\mu}$ and $K_{\mu\nu}$ with the bold ones, i.e., $\vecb{n}_\mu$ and $\vecb{K}_{\mu\nu}$ defined in Section~\ref{sec2:EFT_general}.}
\begin{align}
\bar{T}_{\mu\nu} = & -(\Lambda + c \bar{\vecb g}^{\tau\tau} - \bar{\vecb n}^\lambda \partial_\lambda \tilde{d} + \alpha \bar{\vecb K}^\lambda_{\ \sigma}\bar{\vecb K}^{\sigma}_{\ \lambda} + \xi_1 \bar{\vecb E}^\lambda_{\ \sigma} \bar{\vecb E}^\sigma_{\ \lambda} + \xi_2 \bar{\vecb B}^\lambda_{\ \sigma} \bar{\vecb B}^\sigma_{\ \lambda}) \bar{g}_{\mu\nu} \nonumber \\
&-(2c\bar{\vecb{g}}^{\tau\tau}+\bar{\vecb n}^\lambda \partial_\lambda \tilde{d} - \alpha \bar{\vecb K}^\lambda_{\ \sigma}\bar{\vecb K}^{\sigma}_{\ \lambda})\bar{\vecb{n}}_\mu\bar{\vecb{n}}_\nu
- 2  \bar{\vecb n}_{(\mu}\partial_{\nu)} \tilde{d} + 2 \alpha \bar{\vecb K}^\sigma_{\ \mu} \bar{\vecb K}_{\sigma \nu}
 - 2\bar{\nabla}_\lambda (\alpha \bar{\vecb K}^\lambda_{\ (\mu} \bar{\vecb n}_{\nu)}) \nonumber \\ 
&+ \bar{\nabla}_\lambda (\alpha \bar{\vecb K}_{\mu\nu} \bar{\vecb n}^\lambda) + 2 \xi_1 \bar{\vecb E}_{(\mu| \alpha} \bar{\vecb E}^\alpha_{\ \, |\nu)} + 2 \xi_2 \bar{\vecb B}_{(\mu| \alpha} \bar{\vecb B}^\alpha_{\ \, |\nu)} - 2\xi_1 \bar{\vecb E}^{\alpha\beta} \mathcal{E}_{\alpha\beta \mu\nu} - 2\xi_2 \bar{\vecb B}^{\alpha\beta} \mathcal{B}_{\alpha\beta \mu\nu} \;.  
\label{eq:T_munu_EFT}
\end{align}
Note that, throughout the paper, we use a prime to denote the derivative with respect to $r$.
The tensors~$\mathcal{E}_{\alpha\beta \mu\nu}$ and $\mathcal{B}_{\alpha\beta \mu\nu}$ in (\ref{eq:T_munu_EFT}) are respectively defined as variations of $\vecb{E}_{\mu\nu}$ and $\vecb{B}_{\mu\nu}$ with respect to the metric tensor, i.e., $\delta \vecb{E}_{\mu\nu} \equiv \mathcal{E}_{\mu\nu\alpha\beta } \delta g^{\alpha \beta} $ and $\delta \vecb{B}_{\mu\nu} \equiv \mathcal{B}_{\mu\nu\alpha\beta } \delta g^{\alpha \beta}$,
whose explicit expressions are given by 
    \begin{align}
    \mathcal{E}_{\mu\nu\alpha\beta}
    &=-\frac{1}{2}\bar{\vecb{E}}_{\mu\nu}\bar{g}_{\alpha\beta}+\bar{\vecb{E}}_{\mu\nu}\bar{\vecb{n}}_\alpha\bar{\vecb{n}}_\beta-\bar{\epsilon}_{\mu\nu \sigma (\alpha} \bar{\vecb n}_{\beta )} \bar{\vecb n}^\lambda \bar{F}^\sigma_{\ \lambda} + \bar{\epsilon}_{\mu\nu \sigma (\alpha} \bar{F}_{\beta) \lambda} \bar{\vecb n}^\sigma \bar{\vecb n}^\lambda + \bar{\epsilon}_{\mu\nu \lambda \sigma} \bar{\vecb n}^\lambda \bar{\vecb n}_{(\alpha} \bar{F}^\sigma_{\ \beta)}\;, \\
    \mathcal{B}_{\mu\nu\alpha\beta}
    &=2\bar{\vecb{h}}^\lambda_{\ [\mu}\bar{\vecb{n}}_{\nu]}(\bar{\vecb{n}}_\alpha\bar{\vecb{n}}_\beta\bar{\vecb{n}}^\sigma\bar{F}_{\lambda\sigma}-\bar{\vecb{n}}_{(\alpha}\bar{F}_{\beta)\lambda})\;.
    \end{align}
Note that the results~\eqref{Gmunu-components} and \eqref{eq:T_munu_EFT} can be used for any background configuration of the vector field.

In order to look for possible static and spherically symmetric solutions, we focus on the so-called stealth solutions. These solutions are characterized by those vector field configurations which provide effective background energy-momentum tensor in the form of cosmological constant, i.e., $\bar{T}_{\mu\nu}\propto \bar{g}_{\mu\nu}$. In this regard, we can find background metric solutions with the same form as the general relativity solutions such as the Schwarzschild solution. In order to do so, we consider the following condition:
\begin{align}\label{stealth-X}
\bar{X} = -\frac{1}{2}\bar{A}_\mu \bar{A}^\mu = \frac{q^2}{2} \;,
\end{align}
where $q$ is a constant. Substituting the background metric~\eqref{eq:bg_metric_rt} into the above condition, we find the following background configuration for the gauge field as a simple solution:
\begin{align}\label{eq:gauge_bg}
\bar{A}_t = q \;, \quad \bar{A}_r = \pm q \sqrt{\frac{1 - A}{AB}} \;.
\end{align}
Note that, as pointed out in \cite{Minamitsuji:2021gcq}, the background solutions in this case can be mapped to those of scalar-tensor theories via the transformation~$A_\mu \rightarrow \partial_\mu \Phi$, where $\Phi$ is a scalar field. 
The background configuration~\eqref{eq:gauge_bg} looks much simpler in the Lema\^{\i}tre coordinates. For the plus branch, we find
\begin{align}
\bar{A}_\tau=q\;, \quad
\bar{A}_\rho =0\;,
\label{eq:gauge_bg1}
\end{align}
while we find
$\bar{A}_\tau=q(2 - A)/A$ and $\bar{A}_\rho=-2q(1 - A)/A$ for the minus branch. Comparing \eqref{eq:gauge_bg} and \eqref{eq:gauge_bg1}, we see that vector field has only a non-vanishing temporal component in the coordinate $(\tau,\rho)$ while it has both non-vanishing temporal and spatial components in the coordinate $(t,r)$. Thus, the former coordinate is more convenient than the latter one and that is why we have performed the coordinate transformation \eqref{eq:coord_tranf}. In this regard, we focus on the background configuration defined by the metric~\eqref{eq:metric_bg} and the vector field configuration~\eqref{eq:gauge_bg1} in what follows. Note that this particular configuration was first obtained in \cite{Cheng:2006us} in the context of gauged ghost condensate (see also \cite{Chagoya:2016aar,Minamitsuji:2017aan,Minamitsuji:2021gcq}) and it was shown to be closely related to the stealth solutions in scalar-tensor theories~\cite{{Mukohyama:2005rw,Babichev:2013cya,Kobayashi:2014eva,BenAchour:2018dap,Motohashi:2019sen,Motohashi:2018wdq,Takahashi:2020hso}}.

In general, without assuming any specific form of the background for the gauge field, the equation of motion~(\ref{eq:eins}) is complicated. However, huge simplification happens in the case of the background configuration~\eqref{eq:gauge_bg1} since both $\vecb{E}_{\mu\nu}$ and $\vecb{B}_{\mu\nu}$ vanish at the background level, so that there will be no contributions coming from the $\xi_1(r)$- and $\xi_2(r)$-operators.
Also, we have $\bar{\vecb{n}}_\mu=-\delta^\tau_{\ \mu}/\sqrt{-\bar{g}^{\tau\tau}}$, i.e., $\bar{\vecb{n}}_\mu$ (and hence $\bar{\vecb{K}}^{\mu}_{\ \nu}$) is no longer affected by the gauge field.
Therefore, the Einstein equation~(\ref{eq:eins}) becomes	
\begin{equation}
		\begin{split}
			\Lambda- (1 + g_M q)^2c
			&=M_\star^2(\bar{G}^\tau{}_\rho-\bar{G}^\rho{}_\rho)\;, \\
			\Lambda + (1 + g_M q)^2c 
			&=-M_\star^2\bar{G}^\tau{}_\tau\;, \\
			\left[\partial_\rho\bar{K}+\frac{1-A}{r}\left(\frac{B}{A}\right)'\,\right]\alpha+\frac{A'B}{2A}\alpha'+\sqrt{\frac{B(1 - A)}{A}}\, \tilde{d}'
			&=-M_\star^2\bar{G}^\tau{}_\rho\;, \\
			\frac{1}{2r^2}\sqrt{\frac{B}{A}}\left[r^4\sqrt{\frac{B}{A}}\left(\frac{1-A}{r^2}\right)'\alpha\right]'
			&=M_\star^2(\bar{G}^\rho{}_\rho-\bar{G}^\theta{}_\theta)\;,
		\end{split} \label{EOM_BG}
	\end{equation}
where
    \begin{align}
    \bar{K}\equiv -\frac{2}{r}\sqrt{\frac{B}{A(1-A)}}\bigg(1-A-\frac{rA'}{4}\bigg)\;.
    \end{align}
Notice that, due to the symmetry of the background metric, only the above four components of the Einstein equation are independent. We see that when $g_M = 0$, the background equations above reduce to the ones found in the case of scalar-tensor theories, where the kinetic term of the scalar field is assumed to be constant~\cite{Mukohyama:2022enj,Mukohyama:2022skk,Mukohyama:2023xyf}.
For $g_M \neq 0$, we obtain additional contributions in front of the parameter~$c$ in the first two equations of (\ref{EOM_BG}), whereas the last two equations remain the same as in \cite{Mukohyama:2022enj,Mukohyama:2022skk,Mukohyama:2023xyf}.
These background equations provide relations among the EFT coefficients, given the metric functions~$A(r)$ and $B(r)$, and the gauge-field value~$q$.
In particular, one can obtain conditions under which our EFT admits the Schwarzschild(-de Sitter) solutions as an exact solution with the condition~\eqref{stealth-X}. Indeed, for the metric functions
    \begin{align}
    A(r)=B(r)=1-\frac{r_s}{r}-\frac{\Lambda_{\rm eff}}{3}r^2\;,
    \label{SdS}
    \end{align}
with $r_s$ and $\Lambda_{\rm eff}$ being constants, the background equations~\eqref{EOM_BG} yield the following conditions on the tadpole functions:
\begin{equation}
	\Lambda=M_\star^2 \Lambda_{\rm eff}\;, \qquad
	c=0\;, \qquad
	\alpha'=0\;, \qquad
	\bar{K}'\alpha+\tilde{d}'=0\;.
	\label{eq:existence_stealth}
\end{equation}
Conversely, if one regards the EFT coefficients as an input, then it is possible to fix the functional forms of $A(r)$, $B(r)$, and the constant~$q$ by using the above equations. 
For other background configurations of $A_\mu$ on a static and spherically symmetric metric, 
one would need to take into account the fact that $\bar{\vecb{E}}^\mu_{\ \nu}$ and $\bar{\vecb{B}}^\mu_{\ \nu}$ do not vanish in general and the fact that the gauge field affects $\bar{\vecb{n}}_\mu$ in general, which make the analysis more involved.

One comment here is in order. From \eqref{Gmunu-components} and the first two equations of (\ref{EOM_BG}), we immediately find that $c=0$ happens for any solution with $A(r)=B(r)$. For example, we found $c=0$ in \eqref{eq:existence_stealth} for the Schwarzschild(-de Sitter) metric~\eqref{SdS}, which is also the case in the EFT of scalar-tensor theories~\cite{Mukohyama:2022skk}. However, $c=0$ holds in vector-tensor theories even if $A(r)\neq{B(r)}$. This can be easily deduced from tadpole cancellation condition associated with the temporal component of $A_\mu$ in the tadpole action~(\ref{eq:tad_EFT})\footnote{Note that there is no contribution coming from the operators with $\xi_1(r)$ and $\xi_2(r)$ since the backgrounds for both $\vecb{E}^\mu_{\ \nu}$ and $\vecb{B}^\mu_{\ \nu}$ vanish on the configuration~(\ref{eq:gauge_bg1}).}
\begin{align}\label{eq:tad_con_Sph}
g_M (1 + g_Mq) c(r)  = 0 \;,
\end{align}
which implies $c=0$ as far as $g_M (1 + g_Mq)\neq0$. As we will show in the next Subsection, perturbations of the longitudinal mode of the vector field may become strongly coupled for $c=0$ if one naively ignores the effects of the higher-derivative term. Therefore, such an issue may show up in the EFT of vector-tensor theories for the background configuration~(\ref{eq:bg_metric_rt}) and (\ref{eq:gauge_bg1}) even away from the stealth solutions, which is a crucial difference from the case of scalar-tensor theories.

Before closing this Subsection, let us comment on the other gauge-field tadpole cancellation conditions. Actually, such conditions can be straightforwardly obtained by varying the action~(\ref{eq:tad_EFT}) with respect to the spatial components of $A_\mu$; however, these will not lead to an extra non-trivial condition on the parameter~$c$ since the operator~$\vecb{g}^{\tau\tau}$ does not contain terms that are linear in $\delta A_i$ on the backgrounds~(\ref{eq:bg_metric_rt}) and (\ref{eq:gauge_bg1}). In fact, they provide a consistency check with the tadpole cancellation condition associated with the St\"{u}ckelberg field~$\pi$. More precisely, a combination of the derivatives of the gauge-field tadpole conditions should be consistent with the one of the $\pi$ field.

\subsection{Perturbations in decoupling limit}\label{sec:decoupling_pi}

Here, we will analyze the decoupling-limit action for the longitudinal mode~$\pi$ for the background configuration~(\ref{eq:bg_metric_rt}) and (\ref{eq:gauge_bg1}). As we will see, this analysis reveals a generic behavior of the sound speed at leading order in the presence of the inhomogeneities of black hole background.

In our setup, there are two possible decoupling limits: One is the limit in which the three degrees of freedom of the gauge field are decoupled from gravity (gravity decoupling) and the other is the limit where the longitudinal mode of the gauge field is decoupled from the transverse modes (gauge-field decoupling). In this Subsection, for simplicity, we assume both the decoupling limits, i.e., one describes the dynamics of $\pi$ on fixed backgrounds of both the metric and $A_\mu$, neglecting the couplings with gravity and the transverse modes.

Let us only consider the terms that are leading order in the derivative expansion of the EFT action~\eqref{eq:EFT_GP} and relevant to the longitudinal mode of the gauge field in the decoupling limit:
\begin{align}
S =  \int \mathrm{d}^4x \sqrt{-g}\bigg[\frac{M^2_\star}{2}\vecb{R} - \Lambda(r) - c(r) \vecb{g}^{\tau\tau}  + \frac{1}{2}m_2^4(r) \bigg(\frac{\delta \vecb{g}^{\tau\tau}}{-\bar{\vecb{g}}^{\tau\tau}}\bigg)^2 + \cdots
 \bigg] \;, \label{eq:EFT_decoupling}
\end{align}
where $\cdots$ denotes the higher-derivative operators and those irrelevant to the longitudinal mode.\footnote{The first two terms~$M^2_\star \vecb{R}/2 $ and $\Lambda(r)$ will not affect the perturbation of the longitudinal mode in the decoupling limit but are relevant to the background. Hence, we keep them in \eqref{eq:EFT_decoupling}.} As in the previous Subsections, we work in the Einstein frame where $f = 1$ and assume that the EFT coefficients can be treated as functions of the areal radius~$r= r(\rho - \tau)$, compatible with the background configuration~(\ref{eq:bg_metric_rt}) and (\ref{eq:gauge_bg1}). 

The background equations for the action~\eqref{eq:EFT_decoupling} can be deduced from (\ref{EOM_BG}) as a subset with $\tilde{d}=0=\alpha$. It is straightforward to check that the Schwarzschild(-de Sitter) metric~\eqref{SdS} is the solution in this case. As we have mentioned, the tadpole condition of the gauge field leads to $c=0$. However, we shall keep $c$ for a while to see how the tadpole condition $c=0$ will affect the dynamics of the longitudinal mode.

Here, we assume the condition~$A(r) = B(r)$. Then, the quadratic action for $\pi$ from the action~(\ref{eq:EFT_decoupling}) can be written in the form\footnote{If $A(r) \neq B(r)$, there will be a cross term between $\dot{\pi}$ and $\partial_\rho \pi$ in general, which would require a step of diagonalization to properly define the sound speed.}
    \begin{align}
    \mathcal{L}_2 = p_1 \dot{\pi}^2 - p_2 (\partial_\rho \pi)^2 - p_3 \gamma^{ab}\partial_a\pi\partial_b\pi \;, \label{eq:Lagrangian_pi_decouple}
    \end{align}
up to an irrelevant overall factor, where the last term represents the gradient term associated with the angular derivatives.
The coefficients~$p$'s in (\ref{eq:Lagrangian_pi_decouple}) are defined in terms of the EFT coefficients by
\begin{equation}
\begin{split}
    p_1 \equiv r^2 \sqrt{1 - A}\bigg[c + \frac{2 m_2^4}{(1 + g_M q)^2}\bigg] \;, \quad
    p_2 \equiv \frac{r^2 c}{\sqrt{1 - A}} \;,
   \quad  p_3 \equiv c \,\sqrt{1 - A} \;.
    \end{split} \label{eq:parameter_p}
\end{equation}
From the Lagrangian above, one can obtain both radial and angular sound speeds as
\begin{align}\label{eq:c_rho}
    c_\rho^2 = \frac{\bar{g}_{\rho\rho}}{|\bar{g}_{\tau\tau}|}\frac{p_2}{p_1} = \frac{(1 + g_M q)^2 c }{(1 + g_M q)^2 c + 2 m_2^4}\;, 
\end{align}
and 
\begin{align}\label{eq:c_theta}
     c_\theta^2 &= \frac{r^2}{|\bar{g}_{\tau\tau}|} \frac{p_3}{p_1} = \frac{(1 + g_M q)^2 c}{(1 + g_M q)^2 c + 2m_2^4} \;,
\end{align}
respectively.
It is now evident that both sound speeds are proportional to the parameter~$c$. 
However, as mentioned in Subsection~\ref{sec:gravity}, the tadpole cancellation condition~\eqref{eq:tad_con_Sph} for the gauge field imposes $c=0$, so that the sound speeds in \eqref{eq:c_rho} and \eqref{eq:c_theta} identically vanish.

Thus, the system is infinitely strongly coupled and cannot be treated perturbatively. On the contrary, it is known that the EFT of gauged ghost condensate is weakly coupled all the way up to the cutoff scale set by the value of the background vector field~\cite{Cheng:2006us,Mukohyama:2006mm}, albeit the fact that the symmetry breaking pattern is exactly the same in the Minkowski limit. This is because certain higher-derivative terms, called scordatura terms~\cite{Motohashi:2019ymr,Gorji:2020bfl,Gorji:2021isn,DeFelice:2022xvq}, are taken into account in the EFT of gauged ghost condensate, while they are not included in the simple subclass of the EFT considered in this Subsection. A similar situation happens in the case of cosmological background~\cite{Aoki:2021wew} where the scordatura terms are ignored. It is therefore concluded that this simple subclass including the GP theories as a special case is not able to describe perturbations around the type of stealth solutions considered in this Section. One simply needs to include the scordatura terms, e.g., the $\delta\vecb{K}^2$ term,\footnote{Note that $\delta \vecb{K}^\mu_{\ \nu} \delta \vecb{K}^\nu_{\ \mu}$ can also act as a scordatura term in a similar way, while it changes the speed of gravitational waves~$c_{\rm GW}$. The LIGO/Virgo bound ($|c_{\rm GW}-1| \lesssim 10^{-15}$) puts an upper bound on the coefficient of the operator~$\delta \vecb{K}^\mu_{\ \nu} \delta \vecb{K}^\nu_{\ \mu}$.} that are generally present in the EFT action~(\ref{eq:action_uni2}) [or Eq.~(\ref{eq:action_uni})].

Note that the results could change if we would consider, e.g., a more non-trivial background of the gauge field.
For instance, one could consider the Coulomb-like profile found in \cite{Chagoya:2016aar,Minamitsuji:2016ydr,Heisenberg:2017xda,Heisenberg:2017hwb,Minamitsuji:2017aan,Minamitsuji:2021gcq}. 
Although the analysis would be much more complicated than the one we considered in this paper, we expect that changing the background of $A_\mu$ would lead to an interesting phenomenon regarding the propagation of the scalar perturbation as well as the strong coupling problem. We leave these investigations to future work.

\section{Conclusions} \label{sec:conclusions} 
We have constructed the effective field theory (EFT) of vector-tensor theories on an arbitrary background metric. The construction of the EFT in the unitary gauge is based on the fact that a \textit{timelike} background vector field spontaneously breaks both the time diffeomorphism and $U(1)$ symmetry, while leaving a combination of them invariant as a residual $U(1)$ symmetry \eqref{eq:combined_U(1)_time}, on top of the residual spatial diffeomorphism. In Section~\ref{sec2:EFT_general}, similar to the EFT of vector-tensor theories constructed on a cosmological background~\cite{Aoki:2021wew}, we have provided the EFT action \eqref{eq:action_uni} in the unitary gauge as a spacetime integral of the scalar function of quantities covariant under the residual symmetries, for example, $\vecb{g}^{\tau\tau}$ and $ \vecb{K}$. 
This covariant form of the unitary-gauge EFT can be generally applied to an arbitrary background metric. 
Then, expanding such an EFT action around a generic background geometry in terms of perturbations such as $\delta \vecb{g}^{\tau\tau}$, $\delta \vecb{K}$ and perturbations of other building blocks gives rise to the EFT action for perturbations~\eqref{eq:EFT} or \eqref{eq:EFT_GP} with the coefficients depending on both time and spatial coordinates. 
As pointed out in \cite{Aoki:2021wew}, since in this setup the time coordinate is no longer a building block of the EFT (both $\tau$ and $A_\mu$ transform under the residual $U(1)$ symmetry such that the EFT action remains invariant), one is required to impose a set of consistency relations with respect to the temporal direction in order for the EFT to be invariant under the residual $U(1)$ symmetry~\eqref{eq:combined_U(1)_time}. On top of that, as explained in \cite{Mukohyama:2022enj}, another set of consistency relations with respect to the spatial directions must be imposed on the EFT coefficients. This is because the perturbations we have defined in the unitary gauge do not covariantly transform under the 3d spatial diffeomorphism since their background values explicitly depend on the spatial coordinates. We have found that a consistent way of constructing the EFT of vector-tensor theories on a generic background is to expand the unitary-gauge action around such an inhomogeneous background and to impose the two sets of consistency relations, which ensure that the residual symmetries are preserved at any orders in perturbation.  
Therefore, we have obtained the EFT action of perturbations in the unitary gauge in the context of vector-tensor theories.
This EFT can be indeed generally applied to arbitrary backgrounds of both the metric and the gauge field, with the timelike background vector~$\vecb{n}_\mu$. In Section~\ref{sec:Dictionary}, we have provided a dictionary between our EFT coefficients and those of generalized Proca (GP) theories.
It is important to emphasize that the dictionary we have derived holds for any backgrounds of both the metric and the gauge field. Also, this dictionary enables us to identify a simple subclass of the EFT that includes GP theories as a special case. In Section~\ref{sec4_BG_decoupling}, we have studied the background dynamics, assuming that the background metric is static and spherically symmetric and the background gauge field has a constant temporal component given by Eq.~(\ref{eq:gauge_bg1}), based on the simple subclass that contains GP theories. Specifically, we have arrived at the conclusion that the parameter~$c(r)$ of the EFT action vanishes on this particular choice of backgrounds, which agrees with the result found in the similar EFT constructed on a cosmological background~\cite{Aoki:2021wew}. 
Furthermore, in the case of stealth Schwarzschild(-de Sitter) background, we have analyzed the decoupling-limit action for the longitudinal mode~$\pi$ derived from the leading order EFT action~(\ref{eq:EFT_decoupling}), where the terms linear in the extrinsic curvature and its trace, i.e., $ \tilde{d}(r) \vecb{K}$ and $\alpha(r) \bar{\vecb K}^\mu_{\ \nu} \vecb{K}^\nu_{\ \mu}$, as well as the higher-derivative quadratic operators~$\delta\vecb{K}^2$ and $\delta\vecb{K}^\mu_{\ \nu}\delta\vecb{K}^\nu_{\ \mu}$ are absent.
The analysis shows that the sound speeds~($c_\rho^2$ and $c_\theta^2$) of scalar fluctuations are vanishing everywhere, indicating that the system is infinitely strongly coupled. This is in sharp contrast to the case of gauged ghost condensate, in which perturbations are weakly coupled thanks to certain higher-derivative terms, i.e., the scordatura terms. This implies that, in order to describe this type of stealth solutions within the EFT, the scordatura terms must necessarily be taken into account in addition to those already included in the simple subclass.

There are several future directions we would like to investigate further. As already pointed out at the end of Subsection~\ref{sec:decoupling_pi}, it is worth studying the decoupling limit of $\pi$ on other backgrounds of $g_{\mu\nu}$ or $A_\mu$. This would indeed shed light on whether or not the strong coupling we have found in the absence of scordatura terms is also present in other cases. 
Another interesting prospect is to use our EFT to study the dynamics of both tensor and vector perturbations at least at quadratic order in the odd- and even-parity sectors. Especially, since the strong coupling problem we have found in Subsection~\ref{sec:decoupling_pi} is expected to be absent in the odd sector as the longitudinal mode of the gauge field belongs to the even sector, the odd-sector analysis should be straightforwardly carried out. Finally, it would be phenomenologically interesting to use our EFT to study the quasinormal mode spectrum of perturbations as done in \cite{Mukohyama:2023xyf,Konoplya:2023ppx} for the case of EFT of scalar-tensor theories, and to compute the tidal Love number of black holes. 

\section*{Acknowledgements}
This work was supported by World Premier International Research Center Initiative (WPI), MEXT, Japan. The work of K.~A.~was supported by JSPS KAKENHI Grant No.\ JP20K14468. The work of M.~A.~G.~was supported by Mar\'{i}a Zambrano fellowship. The work of K.~T.~was supported by JSPS KAKENHI Grant Nos.\ JP22KJ1646 and JP23K13101. The work of V.~Y.~was supported by JSPS KAKENHI Grant No.\ JP22K20367. M.~A.~G.~and S.~M.~thank organizers of the workshop “CAS-JSPS-IBS CTPU-CGA workshop in cosmology, gravitation and particle physics” in Prague where this work was in its final stage. S.~M.~is grateful for the hospitality of Perimeter Institute, the cosmology group at Simon Fraser University and the Theoretical Physics Institute at University of Alberta, where part of this work was carried out. Research at Perimeter Institute is supported in part by the Government of Canada through the Department of Innovation, Science and Economic Development and by the Province of Ontario through the Ministry of Colleges and Universities. 


\appendix

\section{Mapping between different bases of EFT building blocks}\label{app}

In the main text, we chose $\{\vecb{K}^\mu_{\ \nu},{}^{(3)}\!\vecb{R}^\mu_{\ \nu}\}$ as a part of the EFT building blocks and we did not separate these tensors into the trace and traceless parts as in \eqref{K_R3_decomposition}.
In this Appendix, we comment on the mapping of the EFT to the basis of $\{\vecb{\sigma}^\mu_{\ \nu},\vecb{r}^\mu_{\ \nu},\vecb{K},{}^{(3)}\!\vecb{R}\}$.

Let us first consider the following Lagrangian:
    \begin{align}
    {\cal L}=\vecb{K}^\mu_{\ \nu}\vecb{K}^\nu_{\ \mu}\;,
    \label{Lag_simple_Kmn}
    \end{align}
which can be expressed in terms of $\vecb{\sigma}^\mu_{\ \nu}$ and $\vecb{K}$ as
    \begin{align}
    {\cal L}=\vecb{\sigma}^\mu_{\ \nu}\vecb{\sigma}^\nu_{\ \mu}+\frac{\vecb{K}^2}{3}\;.
    \label{Lag_simple_sigma}
    \end{align}
We now expand the Lagrangian about the background.
In terms of the perturbations defined as
    \begin{align}
    \delta \vecb{K}^\mu_{\ \nu}\equiv \vecb{K}^\mu_{\ \nu}-\bar{\vecb{K}}^\mu_{\ \nu}, \qquad
    \delta\vecb{\sigma}^\mu_{\ \nu}\equiv \vecb{\sigma}^\mu_{\ \nu}-\bar{\vecb{\sigma}}^\mu_{\ \nu}, \qquad
    \delta \vecb{K}\equiv \vecb{K}-\bar{\vecb{K}}\;,
    \end{align}
we obtain
    \begin{align}
    {\cal L}
    =\bar{\vecb{K}}^\mu_{\ \nu}\bar{\vecb{K}}^\nu_{\ \mu}+2\bar{\vecb{K}}^\mu_{\ \nu}\delta\vecb{K}^\nu_{\ \mu}+\delta\vecb{K}^\mu_{\ \nu}\delta\vecb{K}^\nu_{\ \mu}\;,
    \label{Lag_simple_pert_Kmn}
    \end{align}
or
    \begin{align}
    {\cal L}
    =\bar{\vecb{\sigma}}^\mu_{\ \nu}\bar{\vecb{\sigma}}^\nu_{\ \mu}+2\bar{\vecb{\sigma}}^\mu_{\ \nu}\delta\vecb{\sigma}^\nu_{\ \mu}+\delta\vecb{\sigma}^\mu_{\ \nu}\delta\vecb{\sigma}^\nu_{\ \mu}
    +\frac{1}{3}\left(\bar{\vecb{K}}^2+2\bar{\vecb{K}}\delta\vecb{K}+\delta\vecb{K}^2\right),
    \label{Lag_simple_pert_sigma}
    \end{align}
depending on if we choose $\vecb{K}^\mu_{\ \nu}$ or $\{\vecb{\sigma}^\mu_{\ \nu},\vecb{K}\}$ as independent building blocks.
Note that the perturbation of the projection tensor, $\delta \vecb{h}^\mu_{\ \nu}\equiv \vecb{h}^\mu_{\ \nu}-\bar{\vecb{h}}^\mu_{\ \nu}$, does not show up in either \eqref{Lag_simple_pert_Kmn} or \eqref{Lag_simple_pert_sigma}.
This is because $\vecb{h}^\mu_{\ \nu}$ does not appear explicitly in the covariant Lagrangian~\eqref{Lag_simple_Kmn} or \eqref{Lag_simple_sigma}.
However, the situation becomes non-trivial when we try to directly find the mapping between the Lagrangians~\eqref{Lag_simple_pert_Kmn} and \eqref{Lag_simple_pert_sigma} which are expressed in terms of perturbations.
Actually, since
    \begin{align}
    \delta\vecb{K}^\mu_{\ \nu}
    =\delta \vecb{\sigma}^\mu_{\ \nu}+\frac{\delta \vecb{K}}{3}\bar{\vecb{h}}^\mu_{\ \nu}+\frac{\vecb{K}}{3}\delta \vecb{h}^\mu_{\ \nu}\;,
    \end{align}
the last two terms in \eqref{Lag_simple_pert_Kmn} can be rewritten as
    \begin{align}
    \begin{split}
    \bar{\vecb{K}}^\mu_{\ \nu}\delta\vecb{K}^\nu_{\ \mu}
    &=\left(\bar{\vecb{\sigma}}^\mu_{\ \nu}+\frac{\bar{\vecb{K}}}{3}\bar{\vecb{h}}^\mu_{\ \nu}\right)\delta\vecb{\sigma}^\nu_{\ \mu}+\frac{\bar{\vecb{K}}}{3}\delta\vecb{K}+\frac{\vecb{K}}{3}\bar{\vecb{\sigma}}^\mu_{\ \nu}\delta\vecb{h}^\nu_{\ \mu}+\frac{\bar{\vecb{K}}\vecb{K}}{9}\bar{\vecb{h}}^\mu_{\ \nu}\delta\vecb{h}^\nu_{\ \mu}\;, \\
    \delta\vecb{K}^\mu_{\ \nu}\delta\vecb{K}^\nu_{\ \mu}
    &=\delta \vecb{\sigma}^\mu_{\ \nu}\delta \vecb{\sigma}^\nu_{\ \mu}+\frac{\delta \vecb{K}^2}{3}+\frac{2}{3}\bar{\vecb{h}}^\mu_{\ \nu}\delta\vecb{K}\delta\vecb{\sigma}^\nu_{\ \mu}+\frac{2}{3}\vecb{K}\left(\delta \vecb{\sigma}^\mu_{\ \nu}+\frac{\delta \vecb{K}}{3}\bar{\vecb{h}}^\mu_{\ \nu}\right)\delta\vecb{h}^\nu_{\ \mu} \\
    &\quad +\frac{\vecb{K}^2}{9}\delta\vecb{h}^\mu_{\ \nu}\delta\vecb{h}^\nu_{\ \mu}\;,
    \end{split}
    \label{unwanted_delta-h}
    \end{align}
respectively, which involve $\delta\vecb{h}^\mu_{\ \nu}$ explicitly.
Nevertheless, the unwanted $\delta\vecb{h}^\mu_{\ \nu}$'s can be removed from \eqref{Lag_simple_pert_Kmn}.
The point is that the above two terms show up in \eqref{Lag_simple_pert_Kmn} in the combination~$2\bar{\vecb{K}}^\mu_{\ \nu}\delta\vecb{K}^\nu_{\ \mu}+\delta\vecb{K}^\mu_{\ \nu}\delta\vecb{K}^\nu_{\ \mu}$ thanks to the fact that they are derived from the covariant Lagrangian~\eqref{Lag_simple_Kmn}.
Note that this is nothing but the consistency relation we have discussed in Subsection~\ref{sec:expand_EFT} for a more general Lagrangian.
Indeed, starting from \eqref{unwanted_delta-h}, we obtain
    \begin{align}
    2\bar{\vecb{K}}^\mu_{\ \nu}\delta\vecb{K}^\nu_{\ \mu}+\delta\vecb{K}^\mu_{\ \nu}\delta\vecb{K}^\nu_{\ \mu}
    =2\bar{\vecb{\sigma}}^\mu_{\ \nu}\delta\vecb{\sigma}^\nu_{\ \mu}+\delta\vecb{\sigma}^\mu_{\ \nu}\delta\vecb{\sigma}^\nu_{\ \mu}
    +\frac{1}{3}\left(2\bar{\vecb{K}}\delta\vecb{K}+\delta\vecb{K}^2\right)\;,
    \end{align}
where we have used the following two identities to remove $\delta\vecb{h}^\mu_{\ \nu}\delta\vecb{\sigma}^\nu_{\ \mu}$ and $\delta\vecb{h}^\mu_{\ \nu}\delta\vecb{h}^\nu_{\ \mu}$:
    \begin{align}
    \begin{split}
    0&=\vecb{h}^\mu_{\ \nu}\vecb{\sigma}^\nu_{\ \mu}-\bar{\vecb{h}}^\mu_{\ \nu}\bar{\vecb{\sigma}}^\nu_{\ \mu}
    =\bar{\vecb{h}}^\mu_{\ \nu}\delta\vecb{\sigma}^\nu_{\ \mu}+\bar{\vecb{\sigma}}^\mu_{\ \nu}\delta\vecb{h}^\nu_{\ \mu}+\delta\vecb{h}^\mu_{\ \nu}\delta\vecb{\sigma}^\nu_{\ \mu}\;, \\
    0&=\vecb{h}^\mu_{\ \nu}\vecb{h}^\nu_{\ \mu}-\bar{\vecb{h}}^\mu_{\ \nu}\bar{\vecb{h}}^\nu_{\ \mu}
    =2\bar{\vecb{h}}^\mu_{\ \nu}\delta\vecb{h}^\nu_{\ \mu}+\delta\vecb{h}^\mu_{\ \nu}\delta\vecb{h}^\nu_{\ \mu}\;.
    \end{split}
    \end{align}
Now, as expected, the perturbation of the projection tensor has been canceled out and we have recovered (the perturbative part of) Eq.~\eqref{Lag_simple_pert_sigma}.

The same thing should happen for a more general EFT action.
Namely, if we start from our EFT action~\eqref{eq:EFT_GP} and rewrite it in terms of $\{\vecb{\sigma}^\mu_{\ \nu},\vecb{r}^\mu_{\ \nu},\vecb{K},{}^{(3)}\!\vecb{R}\}$, then the perturbation of the projection tensor should be canceled out thanks to the consistency relations.
Therefore, one could choose either $\{\vecb{K}^\mu_{\ \nu},{}^{(3)}\!\vecb{R}^\mu_{\ \nu}\}$ or $\{\vecb{\sigma}^\mu_{\ \nu},\vecb{r}^\mu_{\ \nu},\vecb{K},{}^{(3)}\!\vecb{R}\}$ as a part of the EFT building blocks, and there exists a one-to-one correspondence between the EFT actions written in the two different manners.

{}
\bibliographystyle{utphys}
\bibliography{bib_v5}

\end{document}